\documentclass{iopart}

\pdfoutput=1

\newcommand{\roughly}{\mathchar"5218\relax} 

\usepackage{graphicx}
\usepackage{latexsym}
\usepackage{iopams, amsopn, amstext, wasysym}
\usepackage{colonequals}
\usepackage{dcolumn}
\usepackage{setspace}
\usepackage{subfigure}
\usepackage{bm}
\usepackage{url}
\usepackage{citesort}
\usepackage{rotating}
\usepackage{braket}
\usepackage{units} 
\usepackage[usenames,dvipsnames]{color}

\usepackage{ulem}
\newcommand{\Ytwo}{{{}^{-2}Y}}

\newcommand{\define}{\colonequals}

\def\IL{\relax{\rm I\kern-.18em L}}

\newcommand{\Ms}{\ensuremath{\mathrm{M}_{\odot}}}
\newcommand{\msun}{\ensuremath{\mathrm{M}_\odot}}

\newcommand{\Overlap}{\Braket}

\def\ltsima{$\; \buildrel < \over \sim \;$}
\def\simlt{\lower.5ex\hbox{\ltsima}}
\def\gtsima{$\; \buildrel > \over \sim \;$}
\def\simgt{\lower.5ex\hbox{\gtsima}}

\makeatletter
\let\protect\relax
{\catcode`\|=\active
  \xdef\InnerProduct{\protect\expandafter\noexpand\csname InnerProduct
\endcsname}
  \expandafter\gdef\csname InnerProduct \endcsname#1{%
    \begingroup
    \ifx\SavedDoubleVert\relax
    \let\SavedDoubleVert\|\let\|\IpDoubleVert
    \fi
    \mathcode`\|32768\let|\IPVert
    \left({#1}\right)
    \endgroup
  }
}
\def\IPVert{\@ifnextchar|{\|\@gobble}
     {\egroup\,\mid@vertical\,\bgroup}}
\def\IPDoubleVert{\egroup\,\mid@dblvertical\,\bgroup}
\let\SavedDoubleVert\relax
\def\midvert{\egroup\mid\bgroup}
\def\SetVert{\@ifnextchar|{\|\@gobble}
    {\egroup\;\mid@vertical\;\bgroup}}
\def\SetDoubleVert{\egroup\;\mid@dblvertical\;\bgroup}
\def\mid@vertical{\mskip1mu\vrule\mskip1mu}
\def\mid@dblvertical{\mskip1mu\vrule\mskip2.5mu\vrule\mskip1mu}
\makeatother

\usepackage{superscriptaddress}

\begin{document}

\title[NINJA-2 waveform catalog]{The NINJA-2 catalog of hybrid
post-Newtonian/numerical-relativity waveforms for non-precessing
black-hole binaries}

\newcommand{\AEIGolm}{\affiliation{Max-Planck-Institut f\"ur
  Gravitationsphysik, Albert-Einstein-Institut, Am M\"uhlenberg 1,
  D-14476 Golm, Germany}} %
\newcommand{\AEIHannover}{\affiliation{Max-Planck-Institut f\"ur
  Gravitationsphysik, Albert-Einstein-Institut, Callinstra{\ss}e 38,
  D-30167 Hannover, Germany}} %
\newcommand{\Birmingham}{\affiliation{School of Physics and Astronomy,
    University of Birmingham, Edgbaston, Birmingham B15 2TT, UK}} %
\newcommand{\CaltechLIGO}{\affiliation{LIGO -- California Institute of
    Technology, Pasadena, California 91125, USA}} %
\newcommand{\CaltechTAPIR}{\affiliation{Theoretical Astrophysics
    130-33, California Institute of Technology, Pasadena, California 91125, USA}} %
\newcommand{\Cambridge}{\affiliation{Department of Applied Mathematics
    and Theoretical Physics, Centre for Mathematical Sciences,
    Wilberforce Road, Cambridge CB3 0WA, UK, and King's College,
    Cambridge CB2 1ST, UK}} %
\newcommand{\Cardiff}{\affiliation{School of Physics and Astronomy,
    Cardiff University, The Parade, Cardiff, UK}} %
\newcommand{\Carleton}{\affiliation{Physics \& Astronomy, Carleton
    College, Northfield, MN, USA}} %
\newcommand{\CITA}{\affiliation{Canadian Institute for Theoretical
    Astrophysics, University of Toronto, 60~St.~George Street,
    Toronto, Ontario M5S 3H8, Canada}} %
\newcommand{\Cornell}{\affiliation{Center for Radiophysics and Space
    Research, Cornell University, Ithaca, New York, 14853, USA}} %
\newcommand{\CSICIEEC}{\affiliation{Institut de Ciencies de l'Espai
    (CSIC-IEEC), Campus UAB, Bellaterra, 08193 Barcelona, Spain}} %
\newcommand{\FAU}{\affiliation{Department of Physics, Florida Atlantic
    University, Boca Raton, Florida 33431}} %
\newcommand{\GATech}{\affiliation{Center for Relativistic Astrophysics
    and School of Physics, Georgia Institute of Technology, Atlanta,
    Georgia 30332, USA}} %
\newcommand{\Goddard}{\affiliation{NASA Goddard Space Flight Center,
    Greenbelt, MD 20771, USA}} %
\newcommand{\Hannover}{\affiliation{Max-Planck-Institut f\"ur
    Gavitationsphysik, Hannover, Germany}} %
\newcommand{\INFN}{\affiliation{INFN-Sezione Firenze/Urbino, I-50019
    Sesto Fiorentino, Italy}} %
\newcommand{\Jena}{\affiliation{Theoretisch Physikalisches Institut,
    Friedrich Schiller Universit\"at, 07743 Jena, Germany}} %
\newcommand{\LSUAstro}{\affiliation{Department of Physics \&
    Astronomy, Louisiana State University, Baton Rouge, Louisiana 70803, USA}} %
\newcommand{\LSUComp}{\affiliation{Center for Computation \&
    Technology, Louisiana State University, Baton Rouge, Louisiana 70803, USA}} %
\newcommand{\Mallorca}{\affiliation{Departament de F\'isica,
    Universitat de les Illes Balears, Palma de Mallorca, Spain}} %
\newcommand{\MarylandComp}{\affiliation{Center for Scientific
    Computation and Mathematical Modeling, University of Maryland,
    4121 CSIC Bldg. 406, College Park, Maryland 20742, USA}} %
\newcommand{\MarylandPhysics}{\affiliation{Maryland Center for
    Fundamental Physics, Department of Physics, University of
    Maryland, College Park, Maryland 20742, USA}} %
\newcommand{\Northwestern}{\affiliation{Department of Physics and
    Astronomy, Northwestern University, Evanston, Illinois, USA}} %
\newcommand{\OleMiss}{\affiliation{Department of Physics and Astronomy,
    The University of Mississippi, University, Mississippi 38677, USA}} %
\newcommand{\PennState}{\affiliation{Center for Gravitational Wave
    Physics, The Pennsylvania State University, University Park, Pennsylvania
    16802, USA}} %
\newcommand{\Potsdam}{\affiliation{Max-Planck-Institut f\"ur
    Gravitationsphysik, Am M\"uhlenberg 1, D-14476 Potsdam,
    Germany}} %
\newcommand{\Princeton}{\affiliation{Department of Physics, Princeton
    University, Princeton, New Jersey 08540, USA}} %
\newcommand{\Rochester}{\affiliation{Center for Computational
    Relativity and Gravitation and School of Mathematical Sciences,
    Rochester Institute of Technology, 85 Lomb Memorial Drive,
    Rochester, New York 14623, USA}} %
\newcommand{\Syracuse}{\affiliation{Department of Physics, Syracuse
    University, Syracuse, New York, 13254, USA}} %
\newcommand{\UIB}{\affiliation{Departament de F\'isica, Universitat de
    les Illes Balears, Crta. Valldemossa km 7.5, E-07122 Palma,
    Spain}} %
\newcommand{\UIUC}{\affiliation{Department of Physics, University of
    Illinois at Urbana-Champaign, Urbana, Illinois 61801, USA}} %
\newcommand{\UMass}{\affiliation{Department of Physics, University of
    Massachusetts, Amherst, Massachusetts 01003, USA}} %
\newcommand{\Urbino}{\affiliation{Istituto di Fisica, Universit\`a di
    Urbino, I-61029 Urbino, Italy}} %
\newcommand{\UTAustin}{\affiliation{University of Texas at Austin,
    Austin, Texas, 78712, USA}} %
\newcommand{\UWM}{\affiliation{University of Wisconsin-Milwaukee,
    P.O.~Box 413, Milwaukee, Wisconsin 53201, USA}} %
\newcommand{\Vienna}{\affiliation{Gravitational Physics, Faculty of
    Physics, University of Vienna, Boltzmanngasse 5, A-1090 Vienna,
    Austria}} %
\newcommand{\ISTCENTRA}{\affiliation{CENTRA, Departamento de F{\'i}sica,
    Instituto Superior T{\'e}cnico, Av.~Rovisco Pais 1, 1049-001 Lisboa,
    Portugal}}

\author{P. Ajith} \CaltechLIGO
\author{Michael Boyle} \Cornell
\author{Duncan A. Brown} \Syracuse
\author{Bernd Br\"ugmann} \Jena
\author{Luisa T.\@ Buchman} \CaltechTAPIR
\author{Laura Cadonati} \UMass
\author{Manuela Campanelli} \Rochester
\author{Tony Chu} \CaltechTAPIR \CITA
\author{Zachariah B.\@ Etienne} \UIUC
\author{Stephen Fairhurst} \Cardiff
\author{Mark Hannam} \Cardiff
\author{James Healy} \GATech
\author{Ian Hinder} \AEIGolm
\author{Sascha Husa} \UIB
\author{Lawrence E.\@ Kidder} \Cornell
\author{Badri Krishnan} \AEIHannover
\author{Pablo Laguna} \GATech
\author{Yuk Tung Liu} \UIUC
\author{Lionel London} \GATech
\author{Carlos O.\@ Lousto} \Rochester
\author{Geoffrey Lovelace} \Cornell
\author{Ilana MacDonald} \CITA
\author{Pedro Marronetti} \FAU
\author{Satya Mohapatra} \UMass
\author{Philipp M\"osta} \AEIGolm
\author{Doreen M\"uller} \Jena
\author{Bruno C.\@ Mundim} \Rochester
\author{Hiroyuki Nakano} \Rochester
\author{Frank Ohme} \AEIGolm
\author{Vasileios Paschalidis} \UIUC
\author{Larne Pekowsky} \Syracuse \GATech
\author{Denis Pollney} \UIB
\author{Harald P.\@ Pfeiffer} \CITA
\author{Marcelo Ponce} \Rochester
\author{Michael P\"urrer} \Vienna
\author{George Reifenberger} \FAU
\author{Christian Reisswig} \CaltechTAPIR
\author{Luc\'ia Santamar\'ia} \CaltechLIGO
\author{Mark A.\@ Scheel} \CaltechTAPIR
\author{Stuart L.\@ Shapiro} \UIUC
\author{Deirdre Shoemaker} \GATech
\author{Carlos F.\@ Sopuerta} \CSICIEEC
\author{Ulrich Sperhake} \CSICIEEC \CaltechTAPIR \OleMiss \ISTCENTRA
\author{B{\'{e}}la Szil{\'{a}}gyi} \CaltechTAPIR
\author{Nicholas W.\@ Taylor} \CaltechTAPIR
\author{Wolfgang Tichy} \FAU
\author{Petr Tsatsin} \FAU
\author{Yosef Zlochower} \Rochester


\printauthorlist

\begin{abstract}
  The Numerical INJection Analysis (NINJA) project is a collaborative
  effort between members of the numerical relativity and gravitational
  wave data analysis communities.  The purpose of NINJA is to study the
  sensitivity of existing gravitational-wave search and
  parameter-estimation algorithms using numerically generated waveforms,
  and to foster closer collaboration between the numerical relativity
  and data analysis communities.  The first NINJA project used only
  a small number of injections of short numerical-relativity waveforms, which 
  limited its ability to draw quantitative
  conclusions.  The goal of the NINJA-2 project is to overcome these limitations
  with long post-Newtonian---numerical relativity hybrid waveforms, 
  large numbers of 
  injections, and the use of real detector data. 
  We report on the submission requirements for the NINJA-2
  project and the construction of the waveform catalog.  Eight numerical
  relativity groups have contributed 63 hybrid waveforms consisting of a
  numerical portion modelling the late inspiral, merger, and ringdown
  stitched to a post-Newtonian portion modelling the early
  inspiral.  We summarize the techniques used by each group in
  constructing their submissions.  We also report on the procedures used
  to validate these submissions, including examination in the time and
  frequency domains and comparisons of waveforms from different groups
  against each other.  These procedures have so far considered only the
  $(\ell,m)=(2,2)$ mode.  Based on these studies we judge that
  the hybrid waveforms are suitable for NINJA-2 studies.
  We note some of the plans for these investigations.
\end{abstract}
  
\maketitle

\section{Introduction}
\label{sec:introduction}
A new generation of laser interferometric gravitational-wave detectors
(Advanced LIGO~\cite{Abbott:2007kv,Shoemaker:aLIGO,2010CQGra..27h4006H},
Advanced Virgo~\cite{Accadia:2011zz,aVIRGO}, and LCGT~\cite{Kuroda2010}) is
presently under construction. These second-generation detectors will have an
order of magnitude increase in sensitivity over first generation instruments
and will be sensitive to a broader range of gravitational-wave frequencies. One
of the most widely anticipated sources for this global network of
observatories is the inspiral, merger and ringdown of a binary containing two
black holes~\cite{thorne.k:1987}. Detection of such a 
\emph{binary black hole coalescence} will
allow astronomers and astrophysicists to directly observe the physics of
black-hole spacetimes and to explore the strong-field conditions of Einstein's
theory of general relativity~\cite{Sathyaprakash:2009xs}.

The ability of gravitational-wave astronomers to use the new generation of
observatories to detect and study binary black hole coalescence depends on the
quality of search and source-parameter measurement algorithms. These algorithms
rely on the physical accuracy of the underlying theoretical waveform models.
Developing and testing the algorithms required to achieve the goals of
gravitational-wave astronomy demands close interaction between the
source-modeling and data-analysis communities. The Numerical INJection
Analysis (NINJA) project was created in 2008 to bring these communities
together and to use the recent advances in numerical relativity
(NR)~\cite{Centrella:2010mx} to test analysis pipelines by adding
physically realistic signals to detector noise in software.  We
describe such additions of signals into noise as ``injections.''

The first NINJA project (NINJA-1)~\cite{Aylott:2009ya} considered a total of
23 numerical waveforms, which were injected into Gaussian noise colored with
the frequency sensitivity of first-generation detectors. These data were
analyzed by nine data-analysis groups using both search and
parameter-estimation algorithms.  However, there were two major limitations to
the NINJA-1 analysis: First, to encourage broad participation, no length or
accuracy requirements were placed on the numerical waveforms. Consequently,
many of these waveforms were too short to inject over an astrophysically
interesting mass range without introducing artifacts into the data. The lowest
mass binary considered in NINJA-1 had a total mass of 35$\Ms$, whereas
the mass of black holes could extend below
$5\Ms$~\cite{2011ApJ...741..103F,0004-637X-725-2-1918}.
In NINJA-1, the waveforms were only inspected for obvious, pathological errors and no cross-checks were performed between the submitted
waveforms; and, therefore it was difficult to assess the physical fidelity of the results.
Second, the NINJA-1 data set contained stationary noise with the simulated
signals already injected into the data.  The data set contained only 126
simulated signals, which precluded detailed statistical studies of the effectiveness of
search and parameter estimation algorithms.  Finally, since the data set
lacked the non-Gaussian noise transients present in real detector data, it was
not possible to fully explore the response of the algorithms in a real search 
scenario.  Despite
these limitations, NINJA-1 successfully removed a number of barriers to
collaboration between the source-modelling and data-analysis communities and
demonstrated where further work is needed. The goal of the second NINJA
project (NINJA-2) is to address the deficiencies of NINJA-1 and to perform a
systematic test of the efficacy of data-analysis pipelines in real-world
situations in preparation for Advanced LIGO and Virgo.


This paper reports on the improvements we have made to the NINJA analysis to
address the first of the limitations described above --- the accuracy of the
numerical waveforms. We present the NINJA-2 waveform catalog and describe the
results of the procedures we have used to validate these data.  NINJA-2 places
requirements on the accuracy and length of the contributed waveforms and we
have performed systematic cross-checks of the submitted waveforms. Each
binary black hole simulation in the NINJA-2 catalog must include at least
five orbits of usable data before merger, i.e., neglecting the initial burst
of junk radiation.  The NR waveform amplitude should be accurate to within
$5\%$, and the phase (as a function of gravitational-wave frequency) should have an accumulated
uncertainty over the entire inspiral, merger and ringdown (of the numerical
simulation), of no more than \unit[0.5]{rad}. We also require that numerical
simulations are ``hybridized'' to post-Newtonian (pN) waveforms so that the
resulting waveforms contain enough cycles to allow injections at $M \geq 10\Ms$ in early
Advanced LIGO data. The continued advances in numerical simulations have also
allowed us to study a somewhat larger region of the signal parameter
space; however we
have restricted our attention to non-precessing binaries for
reasons discussed in Sec.~\ref{sec:overview} below.

\begin{figure}
\centerline{  \includegraphics[width=0.8\linewidth]{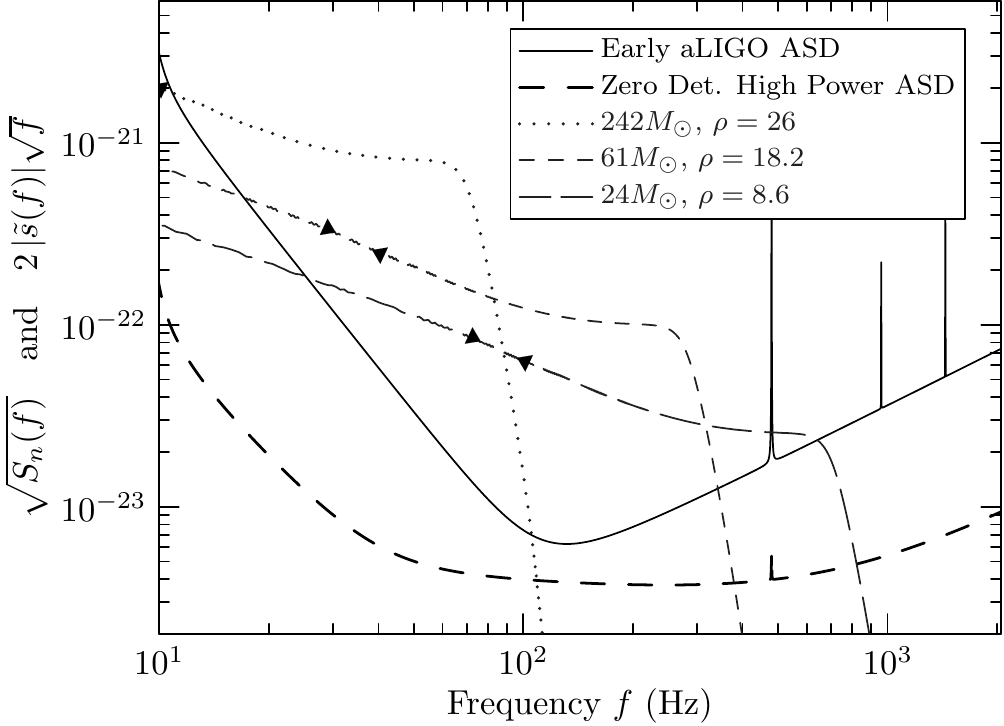}}
  \caption[Hybrid MayaKranc waveform scaled to various total masses]{
  \label{fig:Ninja2StildesAndInitialPSD}
  The hybrid $q=m_1/m_2=2$, non-spinning {\tt MayaKranc} waveform
  scaled to various total masses shown against the early and
  zero-detuned-high-power aLIGO noise curves, shown as amplitude
  spectral densities, the square root of the power spectral densities.
  The triangles represent the starting and ending frequencies of the
  post-Newtonian hybridization region, given in
  table~\ref{tab:ninja2_submissions}.  The total mass of the binary is
  scaled so that the hybridization region ends at \unit[100]{Hz},
  \unit[40]{Hz}, and \unit[10]{Hz}. The amplitude of the signal is
  scaled so that it represents an optimally oriented binary at a
  distance of \unit[1]{Gpc} from the detector. The early aLIGO
  sensitivity is used to compute the signal-to-noise ratio $\rho$.
}
\end{figure}

A subsequent paper will describe the results of using the NINJA-2 waveforms to
study search and parameter estimation algorithms in real detector data. Since
data from the second-generation detectors is not yet available, the NINJA-2
analysis will use data from the first-generation detectors re-colored so that
it has the frequency response expected in the first observing runs of the
Advanced LIGO (aLIGO) detectors --- referred to as ``early
aLIGO''~\cite{G1000176}. Similar noise curves will be used to simulate the
Advanced Virgo detector.  NINJA-2 analyses will use these noise models to
ensure that existing algorithms are optimal when second-generation detectors
come online in $\sim 2015$. The results in this paper use the early aLIGO
sensitivity curve (cf. Fig.~\ref{fig:Ninja2StildesAndInitialPSD}) to study the accuracy of the submitted waveforms.  The
ultimate sensitivity of Advanced LIGO is expected to be significantly better
than this curve. 
To allow the waveforms to be used in
studies using more sensitive noise curves, we have also performed accuracy
studies using the aLIGO \emph{zero-detuned, high-power}~\cite{T0900288}
sensitivity curve. Fig.~\ref{fig:Ninja2StildesAndInitialPSD} shows the two
aLIGO sensitivity curves, characterized by their Amplitude Spectral
Densities (ASD) overlaid with one of the contributed NINJA-2
waveforms. This figure demonstrates that hybridization is necessary to
allow scaling of the numerical waveforms to astrophysically interesting 
masses, and a portion of the present paper studies the hybridization
methods used to construct the NINJA-2 waveforms. 


The remainder of this paper is organized as follows: Section~\ref{sec:overview}
describes in more detail the accuracy requirements that we have placed on
NINJA-2 simulations and presents an overview of the waveform catalog showing
the regions of the binary black hole parameter space covered.
Section~\ref{sec:nrmethods} gives an overview of the numerical methods used to
construct the numerical relativity waveforms and Sec.~\ref{sec:hybridization}
describes the methods that we have used to hybridize the numerical simulations
to pN waveforms. The pN waveforms themselves are summarized in the Appendix. 
Section~\ref{sec:comparisons} describes the methods and results
of the comparisons we have performed between the waveforms.  
Based on these comparisons, we judge the hybrid waveforms suitable for the NINJA-2 project. 
Section~\ref{sec:discussion} summarizes our findings and suggests
directions for future improvements of the catalog, in particular the study of
higher order modes in the waveforms.

\section{Overview of the Waveform Catalog}
\label{sec:overview}

%

Binary black holes formed from the evolution of massive stars are expected to
have circularized before their gravitational-wave frequency reaches the
sensitive band of ground-based detectors, and so we only consider circular
(non-eccentric) binaries. In the NINJA-2 project we further restrict our
attention to binaries that do not undergo precession, i.e., where the spins of
the black holes either vanish or are parallel (or anti-parallel) to the
binary's orbital angular momentum. We do this for two reasons: (i) in trying
to understand the complex phenomenology of the binary parameter space, we
prefer to tackle first a simpler subset, which nonetheless captures the main
features of binary inspiral and merger; (ii) the precessing-binary parameter
space has been sampled by only a handful of numerical simulations. 
The numerical-relativity community is currently exploring the space of precessing
binaries, for example through the numerical relativity-analytical relativity (NRAR) 
project~\cite{ninja-wiki}. Such waveforms
will be used in future NINJA projects that explore precession.
%

\begin{figure}
  \centerline{\includegraphics[width=0.75\linewidth]{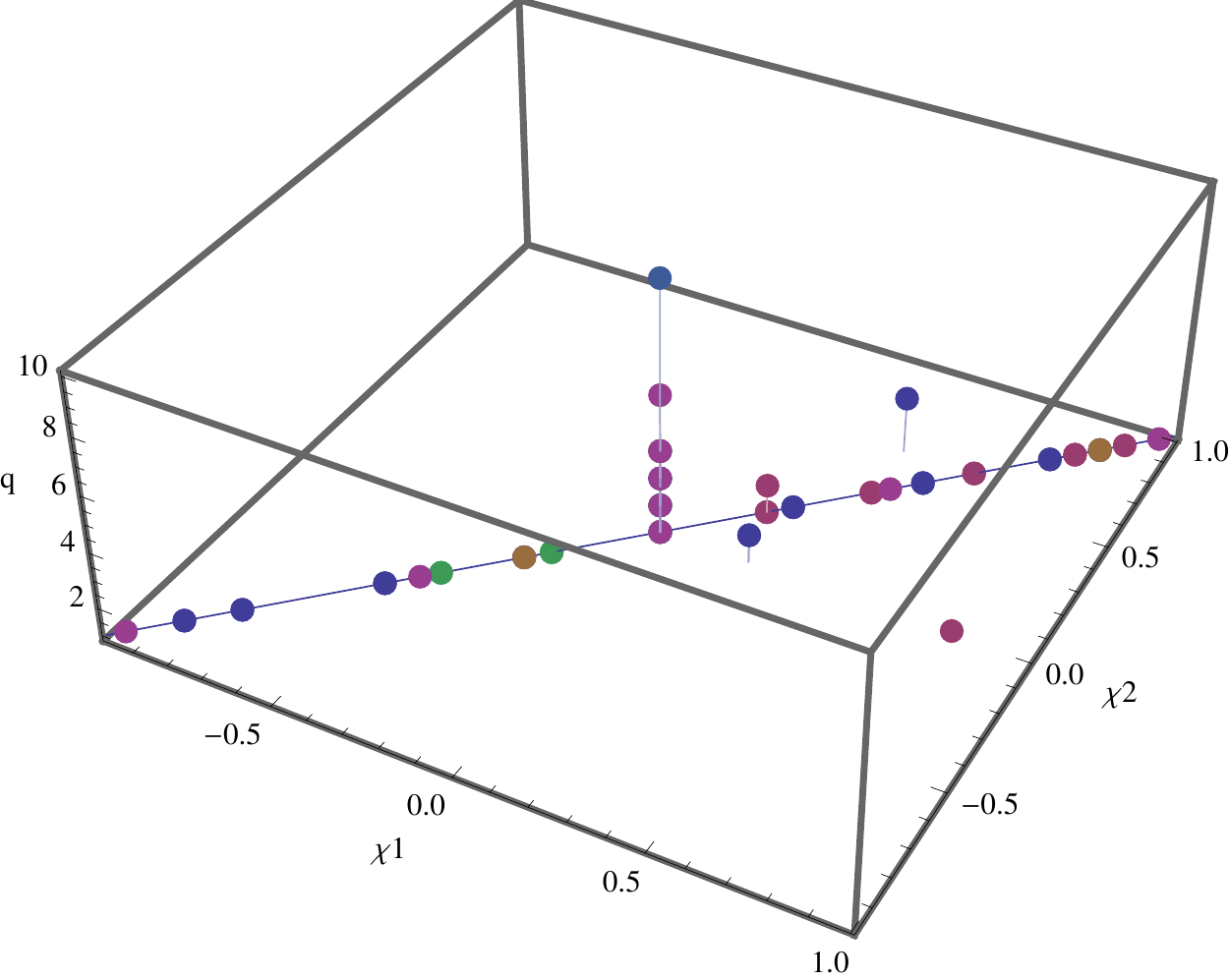}}
  \caption[Parameters of the NINJA-2 submissions]{
  \label{f:ninja2_param_map}
Mass ratio $q$ and dimensionless spins $\chi_i$ of the NINJA-2 hybrid waveform submissions.
}
%
%
%
%
%
%
\end{figure}

The parameters of the black-hole binaries we consider in NINJA-2 are
 the mass of each black hole,
$m_1$ and $m_2$, or equivalently the total mass $M\!=\! m_1 + m_2$ and mass ratio
$q\! =\! m_1/m_2$, and the dimensionless spin-magnitude of each black hole,
$\chi_1\equiv S_1/m_1^2$ and $\chi_2=S_2/m_2^2$. The total mass sets the overall scale of
the system, and can be factored out to leave a three-dimensional parameter space, $\{q,
\chi_1, \chi_2\}$.   Figure~\ref{f:ninja2_param_map} shows the coverage of parameter space (details in Sec.~\ref{sec:nrmethods}).   The sampling is coarse; while the waveforms 
will provide invaluable information within the NINJA-2 project, we expect that
ultimately a more uniform coverage of parameter-space by a much larger number
of configurations will be necessary.

The NINJA-2 requirement of five pre-merger orbits is at the low end of
estimates of sufficient waveform lengths for the construction of accurate
hybrid PN-NR waveforms, as discussed
in~\cite{Hannam:2010ky,Damour:2010zb,MacDonald:2011ne,Boyle:2011dy,Ohme:2011zm}, but we
expect these to be acceptable for the goals of the NINJA-2 project.  The $5\%$
amplitude and \unit[0.5]{rad} phase accuracy requirements were formulated with
typical current waveforms in mind, for example those studied in the Samurai
project~\cite{Hannam:2009hh} and studies performed in preparation for the
NR-AR collaboration project. 
These requirements are consistent for waveforms of similar lengths but may not be
directly applicable to much longer waveforms. For example, in the 25-orbit SpEC
simulations with dimensionless spins $\chi_i=0.97$ the highest- and
second-highest-resolution data differ by roughly \unit[0.6]{rad} at
merger~\cite{Lovelace:2011nu}.  
Note that because this phase-error accumulates over 20 additional inspiral orbits, this waveform would easily satisfy the NINJA-2 phase requirement if it were truncated to minimally meet the NINJA-2 length requirement (although such a truncation would decrease the accuracy of the hybridized waveform).

We require that the hybridized waveforms in the NINJA-2 catalog are long enough to span the
sensitivity bands of the advanced LIGO and Virgo detectors in their early operation.  Specifically, when rescaled  to $10\,\msun$ the hybrids must begin with a gravitational wave frequency of
\unit[20]{Hz} or lower,  i.e. a starting GW frequency of
$M\omega \le 0.006$. This requires extending the NR waveforms to lower
frequencies (i.e. more inspiral cycles) by attaching a pN inspiral waveform
onto the early portion of the NR waveform to produce a \emph{hybrid} pN-NR
waveform. We require that the hybridization be performed at a
gravitational-wave frequency of $M\omega_{22} \leq 0.075$, where
$M\omega_{22}$ is the frequency of the $(\ell, m) = (2, \pm 2)$ harmonic. In
practice, hybridization fits were performed over a frequency range as
summarized in Sec.  \ref{sec:hybridization}, and the average frequency, and
frequency of the average time of the fitting interval were always chosen below
$M\omega_{22} \leq 0.075$, with two exceptions as seen in Table~\ref{tab:ninja2_submissions}: The nonspinning
Llama waveforms at mass ratios $q=1,2$. As seen in Fig.~(\ref{f:ninja2_overlap_test}) these do however show excellent overlaps with comparison waveforms.
%

The waveforms were submitted with the complex GW strain function $h_+-ih_\times$
decomposed into modes using spin-weighted spherical harmonics $\Ytwo_{\ell m}$
of weight $s=-2$.  Although most of the power is in the $(\ell, m)=(2,\pm2)$ modes,
we encouraged (but did not require), the submission of additional subdominant modes. 
The accuracy studies in this paper focus on the
$(\ell, m)=(2,2)$ mode; further work is required to study the
accuracy of the contributed subdominant modes. A total of 63
waveforms from 8 groups were contributed to the NINJA-2 catalog. There are 46
distinct numerical waveforms; some of these waveforms have been hybridized
with multiple pN waveforms.  The NINJA-2 catalog is summarized in
Table~\ref{tab:ninja2_submissions}, and a map of the parameter values is shown
in Fig.~\ref{f:ninja2_param_map}. 
In the next section, we describe in more
detail the numerical methods used to generate these waveforms and present additional
plots in Figs.~\ref{fig:time_plots} 
and~\ref{fig:frequency_sample}.

\begin{table}
  \begin{center}
    \begin{tabular}{|r|r|r|c|r|r|c|c|}
      \hline
      $q$ & $\chi_{1}$ & $\chi_{2}$ & Submission & $1000e$   & $100\,M\omega$ & \# NR & pN \\
      &   &            &            &            & hyb.range & cycles         & Approx \\ 
      \hline
1.0 & -0.95 & -0.95 & SpEC~\cite{Pfeiffer:2002wt,Scheel:2006gg,Lovelace:2011nu,Szilagyi:2009qz,Lovelace:2010ne,Scheel:2008rj,SpECWebsite,Boyle:2007ft,Lindblom:2005qh,Boyle:2009vi} & 1.00 &  3.3 -- 4.1 & 18.42 & T1 \\
1.0 & -0.85 & -0.85 & BAM~\cite{Husa:2007hp,Hannam:2007wf,Ajith:2009bn,Hannam:2010ec,Brugmann:2008zz,Hannam:2007ik} & 2.50 &  4.1 -- 4.7 & 12.09 & T1,T4 \\
1.0 & -0.75 & -0.75 & BAM~\cite{Husa:2007hp,Hannam:2007wf,Ajith:2009bn,Hannam:2010ec,Brugmann:2008zz,Hannam:2007ik} & 1.60 &  4.1 -- 4.7 & 13.42 & T1,T4 \\
1.0 & -0.50 & -0.50 & BAM~\cite{Husa:2007hp,Hannam:2007wf,Ajith:2009bn,Hannam:2010ec,Brugmann:2008zz,Hannam:2007ik} & 2.90 &  4.3 -- 4.7 & 15.12 & T1,T4 \\
1.0 & -0.44 & -0.44 & SpEC~{\footnotesize \cite{Pfeiffer:2002wt,Chu:2009md,Scheel:2006gg,Scheel:2008rj,SpECWebsite,Boyle:2007ft,Lindblom:2005qh}} & 0.04 &  4.3 -- 5.3 & 13.47 & T4 \\
1.0 & -0.40 & -0.40 & Llama~\cite{Pollney:2010hs,Reisswig:2009rx,Pollney:2009yz} &  &  6.1 -- 8.0 & 6.42 & T1,T4 \\
1.0 & -0.25 & -0.25 & BAM~\cite{Husa:2007hp,Hannam:2007wf,Ajith:2009bn,Hannam:2010ec,Brugmann:2008zz,Hannam:2007ik} & 2.50 &  4.5 -- 5.0 & 15.15 & T1,T4 \\
1.0 & -0.20 & -0.20 & Llama~\cite{Pollney:2010hs,Reisswig:2009rx,Pollney:2009yz} &  &  5.7 -- 7.8 & 8.16 & T1,T4 \\
1.0 & 0.00 & 0.00 & BAM~\cite{Husa:2007hp,Hannam:2007wf,Ajith:2009bn,Hannam:2010ec,Brugmann:2008zz,Hannam:2007ik} & 1.80 &  4.6 -- 5.1 & 15.72 & T1,T4 \\
 &  &  & GATech~\cite{Healy:2008js,Healy:2009ir,Bode:2009mt,Herrmann:2007ex,Healy:2009zm,Bode:2011tq,Hinder:2007qu} & 3.00 & 5.5 -- 7.5 & 9.77 & T4 \\
 &  &  & Llama~\cite{Reisswig:2009rx,Pollney:2009yz} &  & 5.7 -- 9.4 & 8.30 & F2 \\
 &  &  & SpEC~\cite{SpECWebsite,Scheel:2006gg,Scheel:2008rj,Pfeiffer:2002wt,Boyle:2007ft,Lindblom:2005qh} & 0.05 & 3.6 -- 4.5 & 22.98 & T4 \\
1.0 & 0.20 & 0.20 & GATech~\cite{Healy:2008js,Healy:2009ir,Bode:2009mt,Herrmann:2007ex,Healy:2009zm,Bode:2011tq,Hinder:2007qu} & 10.00 &  6.0 -- 7.5 & 10.96 & T4 \\
1.0 & 0.25 & 0.25 & BAM~\cite{Husa:2007hp,Hannam:2007wf,Ajith:2009bn,Hannam:2010ec,Brugmann:2008zz,Hannam:2007ik} & 6.10 &  4.6 -- 5.0 & 18.00 & T1,T4 \\
1.0 & 0.40 & 0.40 & GATech~\cite{Healy:2008js,Healy:2009ir,Bode:2009mt,Herrmann:2007ex,Healy:2009zm,Bode:2011tq,Hinder:2007qu} & 10.00 &  5.9 -- 7.5 & 12.31 & T4 \\
 &  &  & Llama~\cite{Pollney:2010hs,Reisswig:2009rx,Pollney:2009yz} &  & 7.8 -- 8.6 & 6.54 & T1,T4 \\
1.0 & 0.44 & 0.44 & SpEC~{\footnotesize \cite{Pfeiffer:2002wt,Chu:2009md,Scheel:2006gg,Scheel:2008rj,SpECWebsite,Boyle:2007ft,Lindblom:2005qh}} & 0.02 &  4.1 -- 5.0 & 22.39 & T4 \\
1.0 & 0.50 & 0.50 & BAM~\cite{Husa:2007hp,Hannam:2007wf,Ajith:2009bn,Hannam:2010ec,Brugmann:2008zz,Hannam:2007ik} & 6.10 &  5.2 -- 5.9 & 15.71 & T1,T4 \\
1.0 & 0.60 & 0.60 & GATech~\cite{Healy:2008js,Healy:2009ir,Bode:2009mt,Herrmann:2007ex,Healy:2009zm,Bode:2011tq,Hinder:2007qu} & 12.00 &  6.0 -- 7.5 & 13.63 & T4 \\
1.0 & 0.75 & 0.75 & BAM~\cite{Husa:2007hp,Hannam:2007wf,Ajith:2009bn,Hannam:2010ec,Brugmann:2008zz,Hannam:2007ik} & 6.00 &  6.0 -- 7.0 & 14.03 & T1,T4 \\
1.0 & 0.80 & 0.00 & GATech~\cite{Healy:2008js,Healy:2009ir,Bode:2009mt,Herrmann:2007ex,Healy:2009zm,Bode:2011tq,Hinder:2007qu} & 13.00 &  5.5 -- 7.5 & 12.26 & T4 \\
1.0 & 0.80 & 0.80 & GATech~\cite{Healy:2008js,Healy:2009ir,Bode:2009mt,Herrmann:2007ex,Healy:2009zm,Bode:2011tq,Hinder:2007qu} & 6.70 &  5.5 -- 7.5 & 15.05 & T4 \\
1.0 & 0.85 & 0.85 & BAM~\cite{Husa:2007hp,Hannam:2007wf,Ajith:2009bn,Hannam:2010ec,Brugmann:2008zz,Hannam:2007ik} & 5.00 &  5.9 -- 6.9 & 15.36 & T1,T4 \\
 &  &  & UIUC~\cite{2009PhRvD..79d4024E} & 20.00 & 5.9 -- 7.0 & 15.02 & T1 \\
1.0 & 0.90 & 0.90 & GATech~\cite{Healy:2008js,Healy:2009ir,Bode:2009mt,Herrmann:2007ex,Healy:2009zm,Bode:2011tq,Hinder:2007qu} & 3.00 &  5.8 -- 7.5 & 15.05 & T4 \\
1.0 & 0.97 & 0.97 & SpEC~{\footnotesize \cite{Pfeiffer:2002wt,Lovelace:2011nu,Scheel:2006gg,Scheel:2008rj,SpECWebsite,Boyle:2007ft,Lindblom:2005qh,Szilagyi:2009qz,Boyle:2009vi}} & 0.60 &  3.2 -- 4.3 & 38.40 & T4 \\
2.0 & 0.00 & 0.00 & BAM~\cite{Husa:2007hp,Hannam:2007wf,Ajith:2009bn,Hannam:2010ec,Brugmann:2008zz,Hannam:2007ik} & 2.30 &  6.3 -- 7.8 & 8.31 & T1,T4 \\
 &  &  & GATech~\cite{Healy:2008js,Healy:2009ir,Bode:2009mt,Herrmann:2007ex,Healy:2009zm,Bode:2011tq,Hinder:2007qu} & 2.50 & 5.5 -- 7.5 & 10.42 & T4 \\
 &  &  & Llama~\cite{Reisswig:2009rx,Pollney:2009yz} &  & 6.3 -- 9.4 & 7.47 & F2 \\
 &  &  & SpEC~{\footnotesize \cite{Pfeiffer:2002wt,Szilagyi:2009qz,Scheel:2006gg,Scheel:2008rj,SpECWebsite,Boyle:2007ft,Lindblom:2005qh,Buchman-etal-in-prep,Boyle:2009vi}} & 0.03 & 3.8 -- 4.7 & 22.34 & T2 \\
2.0 & 0.20 & 0.20 & GATech~\cite{Healy:2008js,Healy:2009ir,Bode:2009mt,Herrmann:2007ex,Healy:2009zm,Bode:2011tq,Hinder:2007qu} & 10.00 &  5.6 -- 7.5 & 11.50 & T4 \\
2.0 & 0.25 & 0.00 & BAM~\cite{Husa:2007hp,Brugmann:2008zz} & 2.00 &  5.0 -- 5.6 & 15.93 & T1,T4 \\
3.0 & 0.00 & 0.00 & BAM~\cite{Husa:2007hp,Hannam:2007wf,Ajith:2009bn,Hannam:2010ec,Brugmann:2008zz,Hannam:2007ik} & 1.60 &  6.0 -- 7.1 & 10.61 & T1,T4 \\
 &  &  & SpEC~{\footnotesize \cite{Pfeiffer:2002wt,Szilagyi:2009qz,Scheel:2006gg,Scheel:2008rj,SpECWebsite,Boyle:2007ft,Lindblom:2005qh,Buchman-etal-in-prep,Boyle:2009vi}} & 0.02 & 4.1 -- 5.2 & 21.80 & T2 \\
3.0 & 0.60 & 0.40 & FAU~\cite{Tichy:2010qa,Bruegmann:2006at,Marronetti:2007ya,Bruegmann:2003aw} & 1.00 &  5.0 -- 5.6 & 18.89 & T4 \\
4.0 & 0.00 & 0.00 & BAM~\cite{Husa:2007hp,Hannam:2007wf,Ajith:2009bn,Hannam:2010ec,Brugmann:2008zz,Hannam:2007ik} & 2.60 &  5.9 -- 6.8 & 12.38 & T1,T4 \\
 &  &  & LEAN~\cite{Sperhake:2006cy,Sperhake:2007gu} & 5.00 & 5.1 -- 5.5 & 17.33 & T1 \\
 &  &  & SpEC~{\footnotesize \cite{Pfeiffer:2002wt,Szilagyi:2009qz,Scheel:2006gg,Scheel:2008rj,SpECWebsite,Boyle:2007ft,Lindblom:2005qh,Buchman-etal-in-prep,Boyle:2009vi}} & 0.03 & 4.4 -- 5.5 & 21.67 & T2 \\
6.0 & 0.00 & 0.00 & SpEC~{\footnotesize \cite{Pfeiffer:2002wt,Szilagyi:2009qz,Scheel:2006gg,Scheel:2008rj,SpECWebsite,Boyle:2007ft,Lindblom:2005qh,Buchman-etal-in-prep,Boyle:2009vi}} & 0.04 &  4.1 -- 4.6 & 33.77 & T1 \\
10.0 & 0.00 & 0.00 & RIT~\cite{Lousto:2010tb,Campanelli:2005dd,Lousto:2010qx} & 0.40 &  7.3 -- 7.4 & 14.44 & T4 \\
      \hline
    \end{tabular}
  \end{center}
\caption[Submissions to NINJA-2]{
\label{tab:ninja2_submissions}
Summary of the NINJA-2 waveform catalog.  Given are mass-ratio $q=m_1/m_2$, 
magnitude of the dimensionless spins $\chi_i=S_i/m_i^2$,
numerical code, orbital eccentricity $e$, 
frequency range of hybridization in $M\omega$,
the number of numerical cycles from the middle of the hybridization
region through the peak amplitude,
and the post-Newtonian Taylor-approximant(s) used for hybridization.
All pN approximants include terms up to 3.5 pN-order, see the
Appendix.
}
\end{table}

\section{Numerical Methods}
\label{sec:nrmethods}

\subsection{Summary of contributions}

The NINJA-2 data set contains both hybrid and original numerical
relativity waveforms, in a data format that is summarized in
Sec.~\ref{sec:dataformat} below, and described in detail in
Ref.~\cite{Brown:2007jx}.  The contributed waveforms cover 29 different
black hole configurations modeling low-eccentricity inspiral, the mass
ratio $q = m_1/m_2$ ranges from 1 to 10, and the simulations cover a
range of non-precessing spin configurations.

The initial frequency $\omega$ of the $(\ell, m) =(2, 2)$ mode for the
numerical waveforms ranges from $0.035/M$ to $0.078/M$, where $M$
denotes the sum of the initial black-hole masses, with both mean and median values of $0.048/M$.
Table~\ref{tab:ninja2_submissions} lists a few key parameters that
distinguish the waveforms, and introduces short tags for the different
contributors:
%
%
\begin{enumerate}
 \item Two groups use versions of the {\tt BAM} code,
  ``BAM''
  labels the Cardiff-Jena-Palma-Vienna collaboration~\cite{Brugmann:2008zz,Husa:2007hp,Hannam:2007ik,Hannam:2007wf,Bruegmann:2003aw}, and
  ``FAU''
  the contribution from the Florida Atlantic group~\cite{Brugmann:2008zz,Husa:2007hp,Tichy:2008du,Bruegmann:2003aw}.

 \item {\tt LazEv} is the RIT
  code~\cite{Zlochower:2005bj,Campanelli:2005dd,Lousto:2007rj}.
 \item {\tt LEAN} has been developed by Ulrich Sperhake~\cite{Sperhake:2006cy,Sperhake:2007gu}.
 \item Two contributions use the {\tt Llama} code~\cite{Reisswig:2009rx,Pollney:2009yz}.  {\tt
    Llama-AEI} is the contribution of the AEI group~\cite{Reisswig:2009rx,Pollney:2009yz}, {\tt
    LLama-PC} is the Palma-Caltech contribution~\cite{Pollney:2010hs}.
 \item GATech is the Georgia Tech group, using the {\tt MayaKranc}
  code~\cite{Vaishnav:2007nm,Hinder:2007qu}.
 \item {\tt SpEC} for the Cornell-Caltech-CITA collaboration
  code~\cite{SpECWebsite,Scheel:2006gg,Boyle:2007ft,Scheel:2008rj},
 \item {\tt UIUC} stands for the University of Illinois at
  Urbana-Champaign team~\cite{Etienne:2008re,Farris:2011vx,ELPS2011}.
\end{enumerate}

The numerical codes follow either of two approaches to solving the
Einstein equations (see \cite{2010nure.book.....B} for a review).  The
{\tt SpEC} code employs the generalized harmonic formulation (see
e.g. Refs.~\cite{Friedrich:2000qv,Pretorius:2004jg}) with gauge
conditions adapted to for black hole
binaries~\cite{Boyle:2007ft,Scheel:2008rj,Szilagyi:2009qz}.  {\tt SpEC} employs black
hole excision to remove singularities in the interiors of the black
holes from the computational domain.  {\tt SpEC}'s initial-data (also using black hole excision)~\cite{Cook:2004kt,Caudill:2006hw,Lovelace:2008tw} is constructed with a pseudo-spectral elliptic solver~\cite{Pfeiffer:2002wt}.

All other codes use the BSSNOK formulation of the Einstein evolution
equations~\cite{Nakamura:1987zz,Shibata:1995we,Baumgarte:1998te} 
with hyperbolic evolution equations for the lapse and
shift in the moving punctures formalism~\cite{Campanelli:2005dd,Baker:2005vv}. 
The 1+log slicing condition for the lapse
function~\cite{Bona:1997hp} is ``singularity-avoiding'': the time
slices freeze in before reaching the singularity in the black hole.
This makes it possible to avoid the use of black hole excision
techniques~\cite{Hannam:2006vv,Hannam:2006xw,Brown:2007tb,Hannam:2008sg},
when evolving the shift vector field $\beta^i$ according to the
$\tilde\Gamma$-driver condition~\cite{Alcubierre:2002kk,vanMeter:2006vi}
(extended to the moving puncture approach which allows for some free
parameters which groups tune individually).


A significant amount of computational infrastructure is shared between
a number of codes.  With the exception of the two groups using {\tt BAM},
all other moving-puncture codes are based on the \texttt{Cactus}
computational toolkit~\cite{Goodale02a,cactus}, the \texttt{Carpet}
mesh-refinement code~\cite{Schnetter:2003rb,carpet} or
the \texttt{EinsteinToolkit} infrastructure~\cite{Loffler:2011ay,einsteintoolkit}. The
\texttt{Cactus}-based codes also use the same apparent horizon finder
code (\textsc{AHFinderDirect})~\cite{Thornburg:2003sf}.
The codes {\tt Llama}, {\tt LazEv}, {\tt Lean} and {\tt MayaKranc} all
use the same pseudospectral solver for the Einstein constraint
equations~\cite{Ansorg:2004ds}, and {\tt BAM} uses a variant
thereof~\cite{Husa:2007hp}.

We will only very briefly summarize the main features of the
numerical methods and codes, as such
information is generally available elsewhere (see references above and the 
NINJA-1 paper~\cite{Aylott:2009ya}).   There are two important exceptions: Two groups use
the new {\tt Llama} code, which is based on a multipatch decomposition of
the numerical grid, and uses spherical coordinates in the outer zones
of the grid, similar to the {\tt SpEC} code.  This allows a more efficient
treatment of the wave zone, and to causally disconnect the outer
boundaries from the wave extraction; in addition {\tt Llama} uses characteristic
extraction to extract the waves directly at null infinity~\cite{Babiuc:2005pg,Babiuc:2008qy}.
The other important new development is the inclusion of two configurations
with BH Kerr parameters $0.95$ and $0.97$~\cite{Lovelace:2010ne,Lovelace:2011nu}, which was made
possible by using superposed Kerr-Schild initial data~\cite{Matzner:1998pt,Lovelace:2008tw} 
in contributions from the {\tt SpEC} collaboration.  Suitably conformally curved initial
data allows BH spins beyond the Kerr parameter of $\approx0.93$, which
is the maximum attainable using conformally flat initial
data~(\cite{Lovelace:2008tw} and the references therein).
For further details on all the numerical codes used, we refer to the
code references listed above, and the recent overview papers on the
numerical solution of the binary black hole 
problem~\cite{2010nure.book.....B,Hinder:2010vn,Pfeiffer:2012}.

\subsection{Data format of contributions}\label{sec:dataformat}

All contributions followed the format specified
in Ref.~\cite{Brown:2007jx}.
%
The data format consists of metadata files and data files with
spherical harmonic modes. The metadata specify the physical parameters
of the BH binaries, such as mass ratio and spins, the initial
frequency of the $(\ell, m)=(2, 2)$ spherical-harmonic mode, 
eccentricity, and also authors, bibliographical references, as well as numerial
methods used.  This metadata format has been significantly extended
since the first NINJA project to contain more information about the
numerical simulations.  For NINJA-1, the waveform data were stored
as 3-column ASCII tables, listing the time at equidistant steps, and
real and imaginary parts of the strain.  For the long hybrid waveforms
in NINJA-2, this format is not efficient; rather we store the time,
amplitude and phase of the modes. The amplitude and phase as functions of time exhibit much less temporal structure than the complex waveform's oscillatory behavior.  Therefore, amplitude and phase at arbitrary time can easily be
recovered by interpolation from a drastically reduced number of time steps. Consequently, data may be provided with unequal time-spacing with only as
many steps as required to accurately regenerate the original hybrid
waveform with simple linear interpolation.

\subsection{Initial data and eccentricity}
\label{ssec:id}

Specifying initial data for black hole binaries in a non-eccentric
inspiral is by itself a non-trivial problem. The elliptic constraint
equations of general relativity need to be solved numerically, and the
free data, which serve as input to these equations and select a
specific configuration of black holes, have to be chosen in a
judicious way (for a general overview see, e.g.,
Ref.~\cite{Cook:2000vr}).


The moving puncture and generalized harmonic codes differ in the way
they specify the free data for the constraint equations, and
correspondingly in how they encode the black hole parameters.  The
codes based on the ``moving puncture'' approach use puncture initial
data~\cite{Bowen:1980yu,Beig:1993gt,Brandt:1997tf,Dain:2001ry} to
model black holes, resulting in initial data that contain a separate
asymptotically flat end within each black hole. The lapse and shift
fields, which determine the coordinate gauge, are initially set to
trivial values, and quickly pick up values that keep the geometry of
the black holes smooth and almost time independent.
The {\tt SpEC} code uses quasi-equilibrium excision initial data where
the interior of the black-hole horizons has been excised from the
numerical grid~\cite{Cook:2004kt,Caudill:2006hw,Pfeiffer:2007yz}. The
constraint equations are solved using the conformal-thin-sandwich
formulation of the initial-value problem~\cite{York:1998hy,Pfeiffer:2002iy}.  
These
data also specify an initial lapse and shift, thus the evolutions can
already be started in an appropriate coordinate gauge.

All of the codes solve the elliptic constraints with pseudo-spectral
numerical codes~\cite{Pfeiffer:2002wt,Ansorg:2004ds}, sharing
numerical infrastructure as noted above.  Initial data always correspond to
the center-of-mass rest frame, such that the net linear momentum vanishes
initially.
All codes take input parameters that determine
the individual black-hole masses $m_i$, spins $\vec{S}_i$, momenta
$\vec{P}_i$ and coordinate separation $D$ of the black holes. We note
however that in the dynamical strong field regime of general
relativity, definitions of mass, spin, and linear momentum
of individual black holes are ambiguous.  
In our case, initial
separations of the black holes are large enough to match with post-Newtonian
waveforms to construct hybrids, and such ambiguities are in some sense hidden in this matching procedure and associated errors.

In order to achieve non-eccentric inspiral, appropriate input
parameters have to be found, i.e., appropriate momenta have to be
chosen for given masses, spins, and separation. No precise definition
of eccentricity is available in general relativity, and one therefore
usually resorts to quantities inspired from Newtonian gravity, which
quantify those oscillations in the black hole tracks or wave signal
that are associated with eccentricity, see e.g., the discussion in
Ref.~\cite{Mroue:2010re}. 
%
The initial parameters are determined by a number of different
methods, using either an initial guess from a standard post-Newtonian
or effective-one-body (EOB) approximation based on Refs.~\cite{Blanchet:2002av,Schafer:2009dq,Damour:2009ic}, or
adding an iterative procedure to further reduce the eccentricity,
based on Refs.~\cite{Boyle:2007ft,Buchman-etal-in-prep,Buonanno:2010yk,Puerrer-etal-in-prep}. 
Measured orbital eccentricities are listed in
Table~\ref{tab:ninja2_submissions}, and have been checked for consistency 
by an independent estimate of the eccentricity based on the 
submitted GW information.

\subsection{Gravitational-wave extraction}
\label{ssec:rad}

The gravitational-wave signal can only be defined unambiguously at
null infinity. The first code that is capable of computing the wave
signal at null infinity for black hole coalescences has become
available since the NINJA-1 project, and is based on combining a
characteristic extraction method with the {\tt Llama} code~\cite{Pollney:2009yz,Reisswig:2009rx}, and
several waveforms have been contributed using this code. Also, these
calculations have provided rigorous error estimates for procedures
where the wave signal is extracted at finite radius, or at a sequence
of radii, combined with an extrapolation to infinite extraction
radius~\cite{BoyleMroue:2009}, 
thus providing justification to such approximations.

Information about extraction radius were included for most contributions.  The extraction radii ranged from $75 M$ to $500 M$, with a median of 
$90 M$. For four contributions the GW signal was read off at null infinity using
characteristic extraction~\cite{Reisswig:2009us,Reisswig:2009rx},
roughly ten contributions extrapolated the GW signal to infinity from a number of extraction radii.
Two techniques are used to extract an approximate gravitational waveform at finite extraction radius: {\tt SpEC} uses the
Regge-Wheeler-Zerilli
method~\cite{ReggeWheeler:1957,Zerilli:1970,Sarbach:2001qq,Rinne:2008vn,BoyleMroue:2009}
to compute the strain directly; all other contributions use the
Newman-Penrose curvature scalar $\psi_4$~\cite{Newman:1961qr}.
Computation of the strain from $\psi_{4}$ requires two time
integrations, which requires the proper choice of the constants of
integration, and may require further ``cleaning procedures'' to get
rid of artifacts resulting from the finite extraction radii, as discussed in detail
in Ref.~\cite{Reisswig:2010di}.  Several techniques were used in practice,
based on time domain~\cite{Damour:2008te}, or frequency domain methods such
as~\cite{Reisswig:2010di}, or heuristic methods of suppressing  low frequency Fourier modes,
or combining the latter with the method of~\cite{Damour:2008te} as in the {\tt BAM} submissions.

\subsection{Numerical Methods, accuracy, and sources of error}
\label{ssec:num}

The numerical algorithms employed in the various codes agree in many
details:  For the time discretization, all
codes use fourth or fifth order accurate explicit algorithms based on
the method of lines. The consistent experience of different groups is
that finite difference errors are dominated by the spatial
discretization, correspondingly all the moving puncture codes use
at least sixth order finite differencing for the
spatial discretization.  The {\tt SpEC} code uses a multi-domain
pseudospectral method with a large number of domains, which shows
exponential convergence.

The moving puncture codes use a simple hierarchy of fixed refinement
boxes which follow the motion of the black holes, and information
between the boxes is communicated using buffer zones and a variant of
Berger-Oliger mesh-refinement.
Interpolation between refinement levels is performed with spatial polynomial interpolation of fifth
({\tt Llama}, {\tt UIUC}, {\tt LazEv}, {\tt Lean} and {\tt MayaKranc}) or
sixth order ({\tt BAM}).  Time interpolation at
mesh-refinement boundaries introduces second-order errors, which does
however not appear to dominate numerical errors.


%


\section{PN-NR Hybrid Waveforms}
\label{sec:hybridization}

While post-Newtonian (pN) methods accurately approximate GW
signals throughout early inspiral, they become increasingly unreliable
towards late inspiral and merger~\cite{Baker:2006ha}.
Numerical relativity (NR) simulations are capable of accurately
computing inspiral, merger, and ringdown portions of binary
coalescence~\cite{Pretorius:2005gq,Campanelli:2005dd,Baker:2005vv,Hinder:2010vn},
but are too computationally expensive to extend far into the inspiral
regime~\cite{Scheel:2008rj}.  Hybrid waveforms are the result of smoothly
blending together pN and NR waveforms to form a waveform that covers
the full BBH dynamics.  For the NINJA-2 project, each NR group has
produced their own hybrid waveform after ensuring that the pN portions
all agree.  We expect some systematic errors resulting from
errors in the pN approximation, the choice of blending region, and the hybridization
method~\cite{Scheel:2008rj,Boyle:2008ge,Santamaria:2010yb,Boyle:2011dy,Hannam:2010ky,MacDonald:2011ne,Ohme:2011zm}.
The NINJA-2 hybrid waveforms do not
contain effective-one-body extensions of pN approximants, although
these have also been used to model complete waveforms (e.g.~\cite{Damour:2007vq,Pan:2011gk,Boyle:2011dy}).

Hybridization uses least-squares fits to
determine the extrinsic parameters for the pN
waveform~\cite{Ajith:2007kx,Boyle:2009dg,Santamaria:2010yb}.  In
general, this is accomplished by evaluating
\begin{eqnarray}
  \label{gen_delta}
  \delta( \vec{u}, a ) = \min_{\{\vec{u},a\}}
  \int^{s_{1}}_{s_{2}}{|\Upsilon_{\text{pN}}(s,\vec{u}) -
    a\Upsilon_{\text{NR}} (s,\vec{u}_{0})|^2\, \rmd s}
\end{eqnarray}
where $\Upsilon$ represents waveform data relating to strain [e.g.,
$h(t)=h_+(t)-i\,h_\times(t)$, $\arg[h(t)]$ or $\tilde{h}(f)$]. If
$\Upsilon$ is derived from the time domain, then $s = t$; if
$\Upsilon$ is in the frequency domain, then $s = f$. For either case,
$[s_1,s_2]$, chosen within the domain of both the pN and NR data sets,
defines the integration interval and, in most cases, the blending
region.  The vector $\vec{u}$ denotes the set of pN--parameters over which
the fitting is performed.  For example, $\vec{u} =
(t_{\text{shift}},\phi_{\text{shift}},\mu)$ corresponds to adjusting time- and phase- shift and the mass ratio of the pN waveform to
match the NR waveform.  The best-fit parameters are denoted
by $\vec{u}^\ast$.  The amplitude scaling factor, $a$, is often 
fixed to $a=1$, but may be included in the fitting 
parameters~\cite{Ajith:2007kx}.  Finally, in the limit
$s_{1} \to s_{2}$, this procedure reduces to enforcing equality of $\Upsilon_{\rm pM}$ and $\Upsilon_{\rm NR}$ at
$s_{1}=s_{2}$, as well as equality of the first derivative.

More explicitly, hybridization may be performed via the following
algorithm:
\begin{enumerate}
  
 \item \label{a1} Choose $[s_1,s_2]$ within the pN and NR data
  sets. Ideally, $[s_1,s_2]$ is sufficiently early so that both pN and
  NR sets should be accurate.
  
 \item \label{a2} Evaluate Eq.~\eref{gen_delta}; apply
  $\{\vec{u}^\ast, a^\ast\}$ to the pN data set, resulting in
  $\Upsilon_{\text{pN}}^\ast$.  Measure error quantities relating to
  fit.
  
 \item \label{a3} If desired, adjust
  $[s_1,s_2]$ and iterate~(\ref{a1}) and~(\ref{a2}), to find a preferred interval $[s_1^\ast,s_2^\ast]$.
  
 \item Defining a monotonic function $z(s)$ such that $z(s<s_{a}) = 0$
  and $z(s>s_{b}) = 1$, the hybrid is given by
  \begin{equation}
    \label{eq:Hybrid}
    \Upsilon_{\text{Hyb}}(s) = [1-z(s)]\, \Upsilon_{\text{pN}}^\ast +
    a^{\ast}\, z(s)\, \Upsilon_{\text{NR}}~.
  \end{equation}
  Note that the transition region $[s_{a}, s_{b}]$ is generally taken
  to be a sub-interval of $[s_1^\ast,s_2^\ast]$, sometimes consisting
  of a single point.
\end{enumerate}

\begin{figure}
  
\includegraphics[width=1\textwidth]{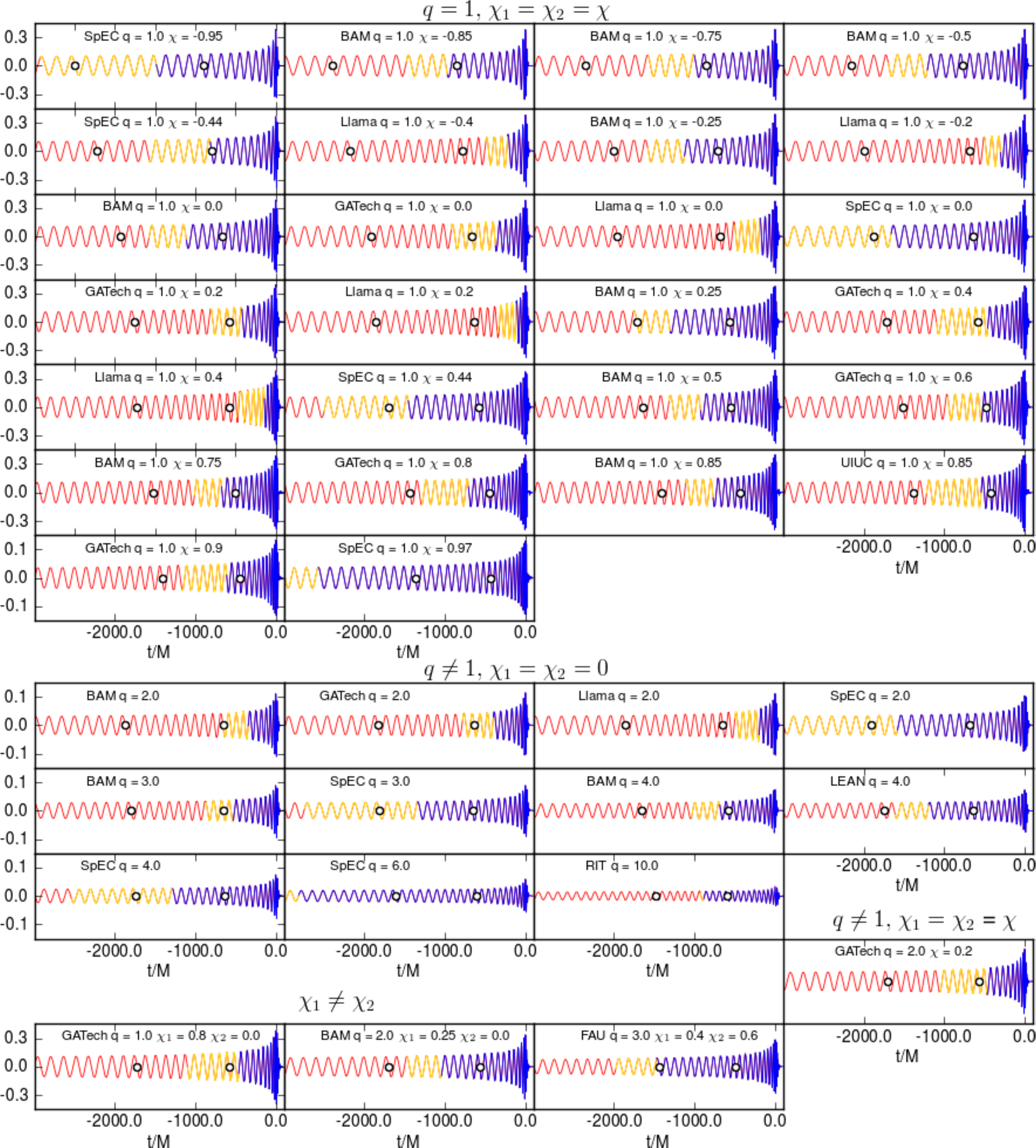}
  \caption{\label{fig:time_plots} {\textbf{Summary of all submitted hybridized waveforms $\mathbf{(r/M) h_{+}}$}} 
  The x-axis shows time in units of M and the y-axis shows the real part of the $(\ell, m) = (2, 2)$ 
  component of the dimensionless wave strain $(r/M) h = (r/M) (h_{+}-ih_{\times})$. 
  The top group shows equal-mass equal-spin waveforms. The middle group shows unequal-mass and zero-spin 
  waveforms, and the bottom group show unequal spin waveforms. The black circles indicate 10 and
  20 GW cycles measured from the waveform peak. 
 The hybridization frequency range is shown in the yellow  
  line. The post-Newtonian part is shown in red line and the NR portion occurring after hybridization 
  is shown in blue line.}
\end{figure}

This general formalism leaves many decisions open, and the following
specific choices were made during hybridization:
\begin{itemize}
\item For the {\tt SpEC} waveforms the integrand of Eq.~(\ref{eq:Hybrid}) was taken 
  to be the
  square of the phase-difference only~\cite{Boyle:2009dg}, $\Upsilon =
  \arg[h(t)]$, and maximation was performed over $t_{\rm shift}$ and
  $\phi_{\rm shift}$ only, without any adjustments to the amplitude
  ($a^*\equiv 1$).  The time-interval $[t_1, t_2]$ was chosen to
  correspond to the frequencies listed in
  Table~\ref{tab:ninja2_submissions}, without any iterative
  adjustments of $t_1$ and $t_2$.
\item The RIT waveforms
similarly use $\Upsilon = \arg[h(t)]$ for Eq.~\eref{gen_delta}, 
but employ the limit as $t_{1} \to t_{2}$ to determine
$\vec{u} = (t_{\text{shift}},\phi_{\text{shift}})$
at $M \omega = 0.075$.
Then, the transition function is $z_{\text{RIT}}(t) = x(t)^3 \,[6 x(t)^2-15
x(t)+10]$, where $x(t) \define (t-t_1)/(t_2-t_1)$ for a finite interval ($t_1 \neq t_2$),
guaranteeing $C^2$ behavior at $t=t_1$ and $t_2$~\cite{Campanelli:2010ac}. 
\item For the {\tt Lean} waveforms hybridization is performed in a similar
  way~\cite{Sperhake:2011zz}, using the transition function
$
  z(t) = 70 x^9 - 315 x^8 + 540 x^7 - 420 x^6 + 126 x^5,
$
and individual mode amplitudes of the PN waveforms are rescaled such
that their average over the matching window agrees with the numerical
result.

\item 
The UIUC waveforms
use $\Upsilon = h(t)$, $a=1$, using as free parameters the initial PN phase and orbital angular frequency. Equation~\eref{eq:Hybrid} is then used to
construct the final hybrid, with $z(t) = (t-t_1)/(t_2-t_1)$ as in \cite{Ajith_2009bn}.
The time-interval $[t_1, t_2]$ was chosen to
correspond to the UIUC frequencies listed in Table~\ref{tab:ninja2_submissions}, 
without any iterative
adjustments of $t_1$ and $t_2$.


\item  The GATech hybridization follows~\cite{Boyle:2009dg} and is done in the time domain with
$\Upsilon = h(t)$ and $\vec{u} = (t_{\text{shift}},
\phi_{\text{shift}})$.  Equation~\eref{gen_delta} is evaluated over
$\{\vec{u},a\}$ and then equation~\eref{eq:Hybrid} is used to
construct the final hybrid, with $z(t) = (t-t_1)/(t_2-t_1)$.  The
fitting intervals are given in Table~\ref{tab:ninja2_submissions}.


\end{itemize}

\begin{figure}
\centerline{ 
\includegraphics[width=0.8\textwidth]{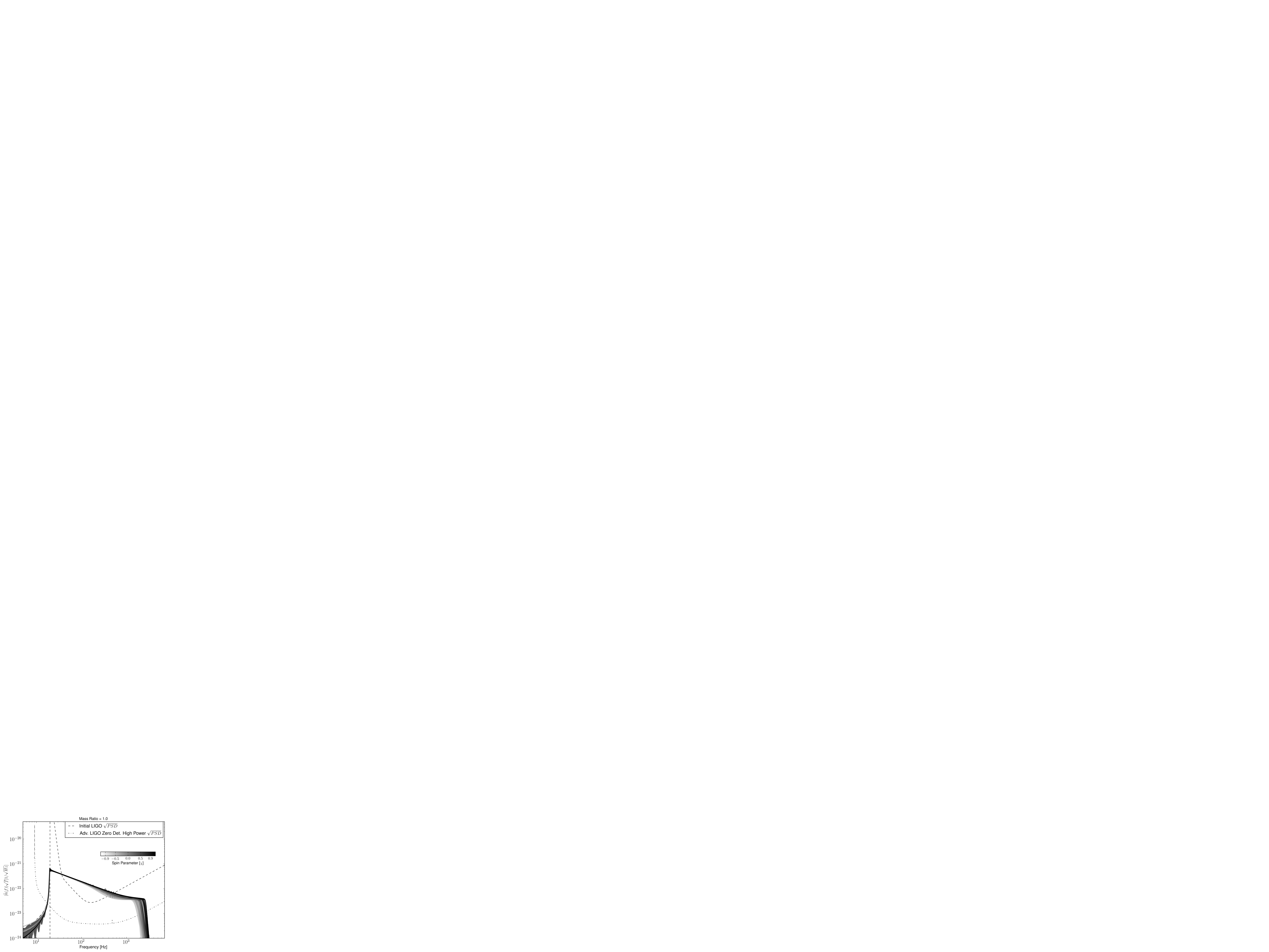}}
 \caption{\label{fig:frequency_sample} {\textbf{Sample frequency
domain plot}} Shown are the plots of $|\tilde{h}(f)|\sqrt{f}$ of $(\ell,m)=(2,2)$
mode for all equal-mass equal-spin waveforms. The
waveforms are scaled to $10\,\msun$ and are placed at 100\,Mpc. 
The Fourier transforms show monotonic behavior in the spin parameter $\chi =
(q\chi_{1}+\chi_{2})/(q+1)$ highlighting the orbital hang-up effect
due to spin. The vertical line indicates 20 Hz, the required upper 
bound on the initial frequency of the hybrids.}
\end{figure}

Figure~\ref{fig:time_plots} and~\ref{fig:frequency_sample} show
exemplary plots of the resulting hybrid waveforms.  For aligned spins,
orbital hangup extends the inspiral to smaller separation and higher
frequencies.  This is apparent in Fig.~\ref{fig:time_plots} in that
the last ten GW cycles (as indicated by the small circles) take less
time, and in Fig.~\ref{fig:frequency_sample} by the shift toward
higher frequencies.

  
%
%
%
%
%


\section{Validation and comparison of hybrid waveforms}
\label{sec:comparisons}
Each NR group verified that their waveforms met the minimum NINJA-2
requirements before submission, as described in
Sec.~\ref{sec:introduction}.  Once submitted, a series of checks were
performed in order to validate the waveforms against each other.  In
the first stage the post-Newtonian expressions and codes were compared
against each other and the literature.  This
resulted in a set of codes in various languages producing waveforms
that agree well in both phase and amplitude (see ~\ref{sec:pn_waveforms}).

\subsection{Time-domain and frequency-domain checks}

In the second stage of validation we examined the $(\ell, m) =(2, 2)$ mode of the hybrid
waveforms.  We first plotted the last 40 cycles of each
waveform --- enough to include the full NR portion, the hybridization
region, and some of the pN portion --- and looked for any anomalies
such as ``kinks'' caused by the hybridization procedures.  
Similar
visual checks were performed on the amplitudes of the Fourier
transforms of the entire waveforms.  This process identified a few
issues, including a bug in one hybridization code, which were then
corrected.  While this was useful, it was only possible due to the
relatively small number of waveforms in NINJA-2 and the fact that only
the $(\ell, m) =(2, 2)$ mode was examined.  For future NINJA projects it will be
necessary to automate this process, the NINJA-2 data analysis may
suggest methods for such automation.

\subsection{Overlap Comparisons}
\label{ssec:ninja2_overlap_comparisons}

In this check the waveforms were compared against each other using
matched-filtering techniques.  The inner product between two real
waveforms $s_1(t)$ and $s_2(t)$ is defined as
\begin{equation}
\label{eq:InnerProduct}
     \InnerProduct{s_1|s_2} 
 = 4\, \Re \int_{0}^\infty df\,
   \frac
     {\tilde{s}_1(f) \tilde{s}_2^\star(f)}
     {S_n(f)}
\end{equation}
where $S_n(f)$ is the power spectral density (see Fig.~\ref{fig:Ninja2StildesAndInitialPSD}).  
The overlap is then
obtained by normalization and maximization over relative time and
phase shifts, $\Delta t$ and $\Delta \phi$.
\begin{equation}
  \label{eq:OverlapDefinition}
  \Overlap{s_1|s_2} \define 
  \max_{\Delta t, \Delta \phi} \frac{\InnerProduct{s_1|s_2}}{
    \sqrt{\InnerProduct{s_1|s_1} \InnerProduct{s_2|s_2}}}.
\end{equation}
There is an important subtlety involved in calculating these overlaps.
Usually, all possible time-shifts are evaluated simultaneously by a
suitable combination of fast Fourier transforms.  This results in
overlaps at time-shifts at discrete values of $\Delta t$ spaced by the
inverse sampling frequency.  The result of time-maximization is then
taken to be the maximum of this discrete time-series.  When comparing
two very similar waveforms, such as two NR simulations of the same
system, the overlap function becomes very sharply peaked.  In units
where $G=c=1$, $1 M_\odot = 4.93\times 10^{-6} s$, and in these units
time shifts of well below $1 M$ can lead to significant changes in the
overlap.  It is therefore imperative that the sample rate used in
calculating the overlap be large enough to find the true maximum.
\begin{figure}
  \includegraphics[width=0.5\linewidth]{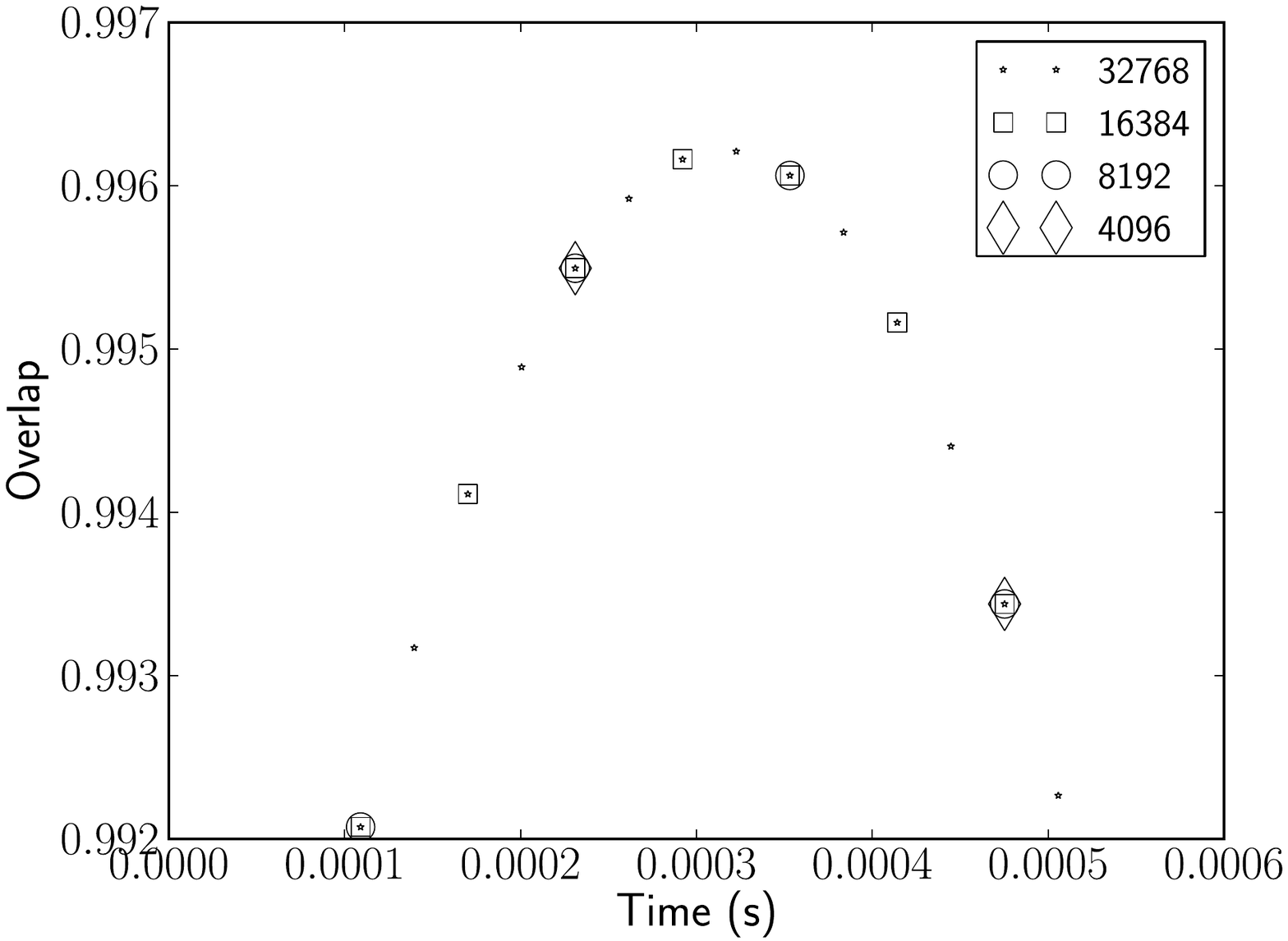}
  \hspace*{-0.05\linewidth}
  \includegraphics[width=0.5\linewidth]{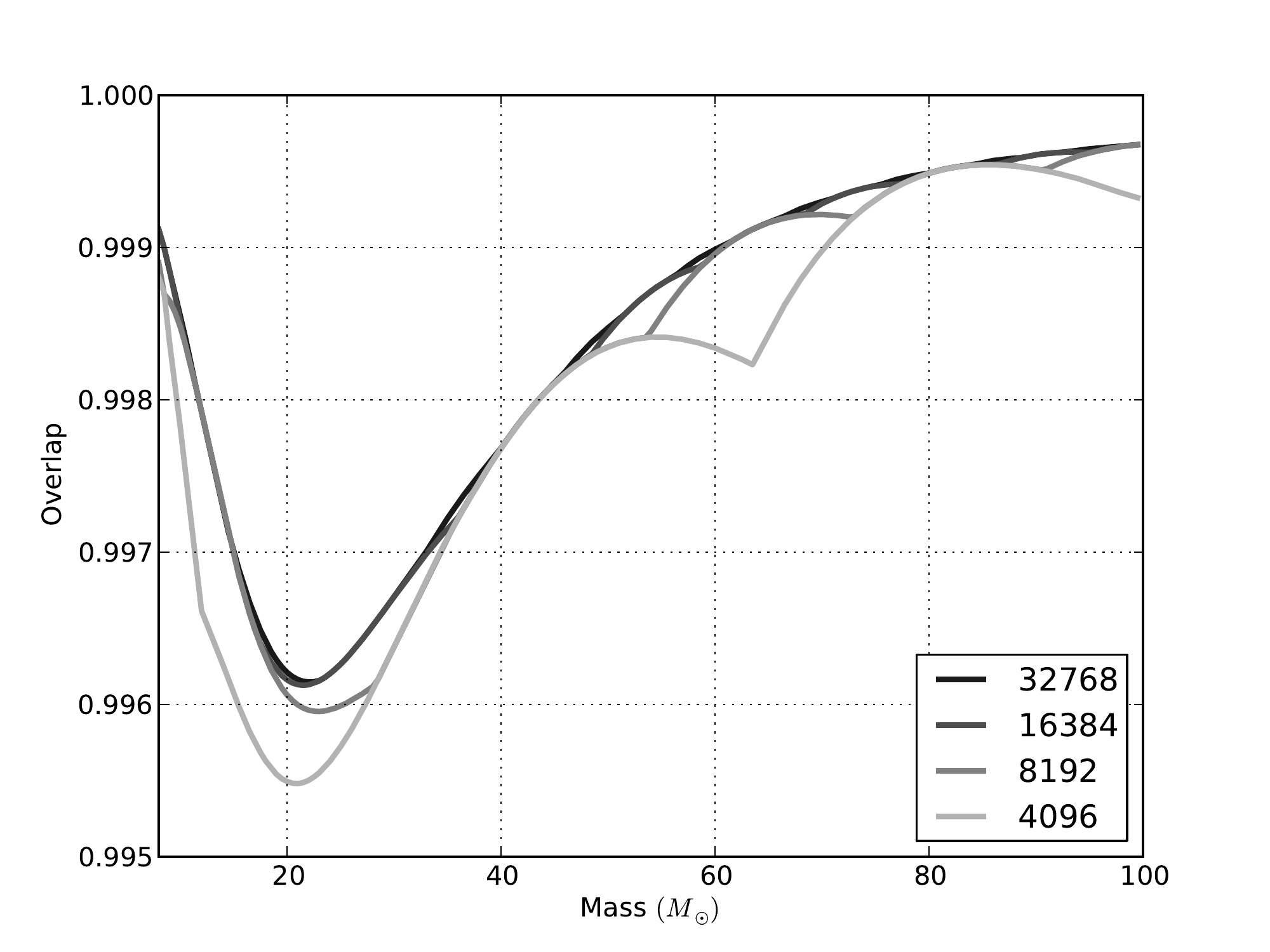}
  \caption[Sensitivity of the overlaps to sample rate]{
  \label{f:overlap_sample_frequency}
  {\bf Left:} Overlap time series between the GATech and {\tt SpEC}
  equal-mass, non-spinning waveforms scaled to $20 M_\odot$ at different
  sample rates.  At \unit[4096]{Hz} the sample point nearest the maximum is
  sufficiently far that the overlap is significantly underestimated, where we 
  are interested in differences to one part in $10^{-5}$.  
  {\bf Right:} Overlaps between the same waveforms as a
  function of mass, at different sample rates. 
  All overlaps are calculated against
  the early aLIGO noise curve.
  }
\end{figure}%

This issue is demonstrated in Fig.~\ref{f:overlap_sample_frequency}.
The left panel shows the overlap function resulting from the
comparison of two equal-mass, non-spinning waveforms sampled at four
different rates.  At \unit[4096]{Hz} the peak of the function is missed and
the overlap is underestimated.  The consequence of this is illustrated
on the right of Fig.~\ref{f:overlap_sample_frequency} which shows that
the overlap as a function of mass exhibits oscillations.  As the mass
changes, the phase-shift $\Delta \phi$ that maximizes
Eq.~(\ref{eq:InnerProduct}) also changes.  For some masses, the
optimal phase-shift will occur at one of the discretely sampled
time-shifts, and the results will be correct.  For other masses, the
true extremum will occur between discretely sampled time-shifts (as
for \unit[4096]{Hz} in the left panel), and the computed overlap will be
erroneously too low.  Henceforth in this paper, all overlaps are
calculated at \unit[32768]{Hz}.  The LIGO/Virgo matched filter searches
operate on data at \unit[4096]{Hz}, however this issue is not a problem in
these searches for reasons that can be seen in these two images.  The
loss of overlap by undersampling is no larger than 0.2\%.  The search
utilizes a bank of templates which discretizes the parameter space.
In constructing the template bank we have already incurred a potential
loss of SNR of 3\%, and this effect is therefore negligible.

\begin{figure}
\centerline{  \includegraphics[width=0.55\linewidth]{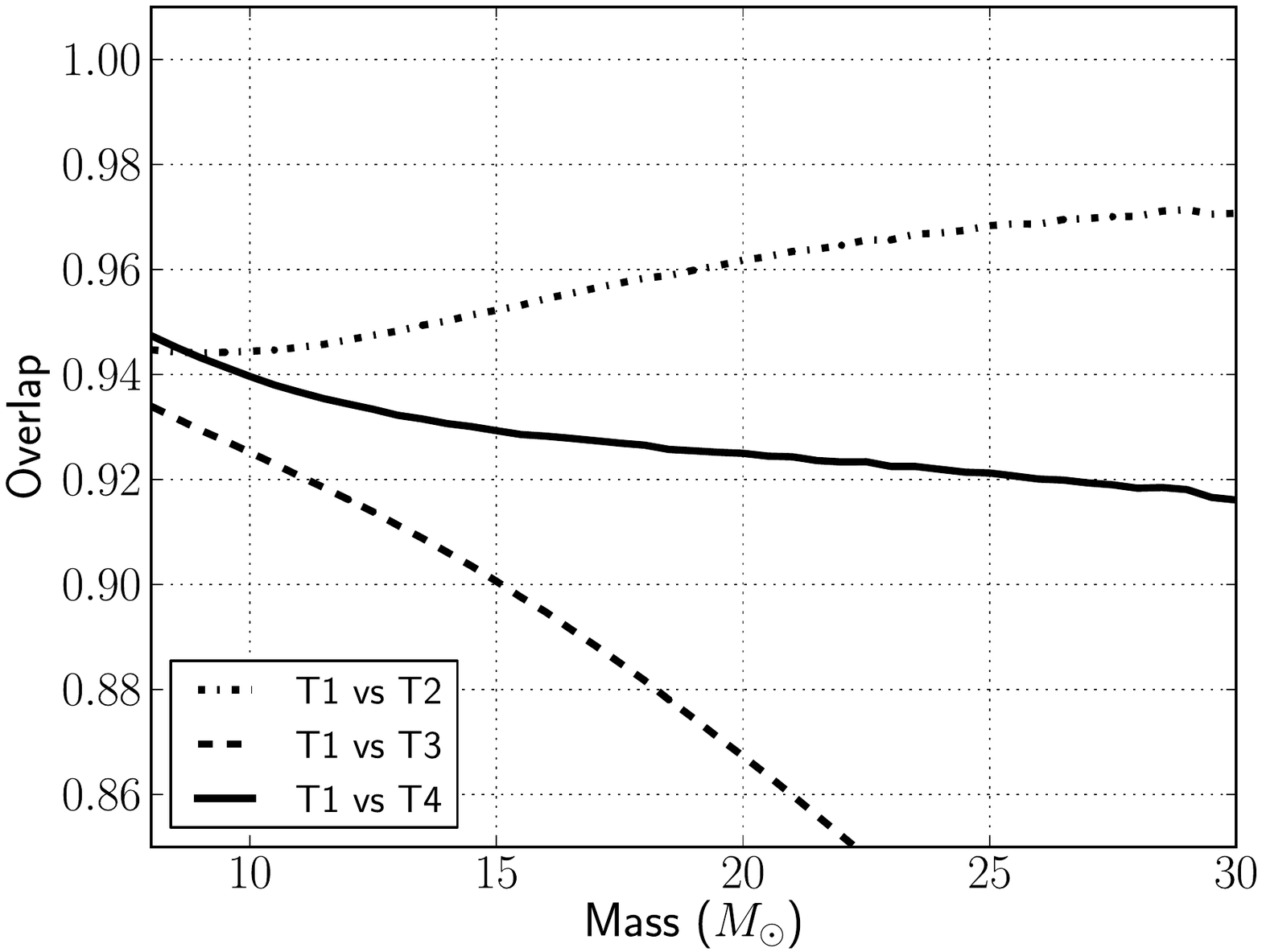}
  \hspace*{-0.05\linewidth}
  \includegraphics[width=0.55\linewidth]{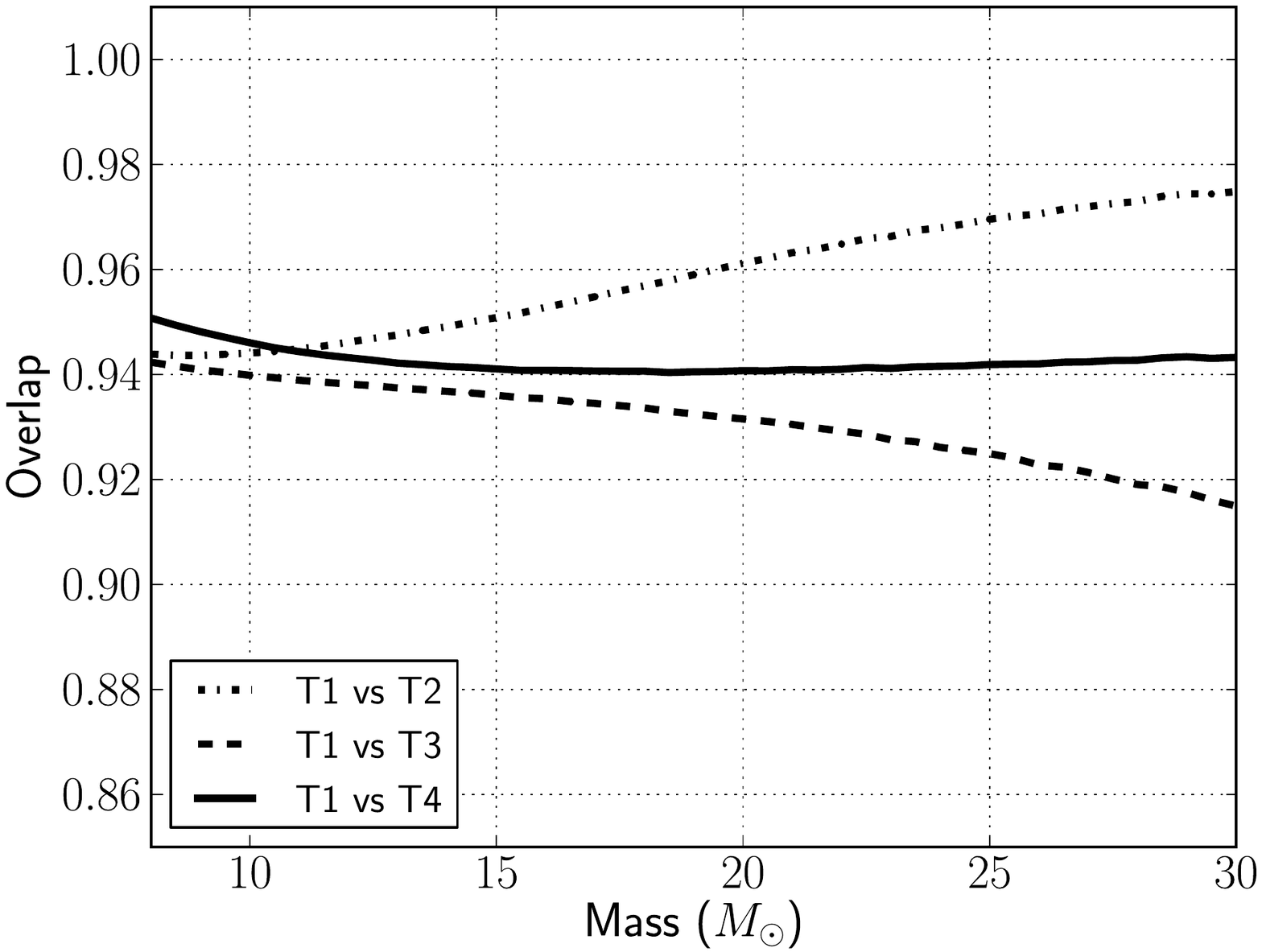}}
  \caption[Overlaps between post-Newtonian waveforms as a function of mass]{
  \label{f:pn_overlaps}
  Overlaps between equal-mass, non-spinning, post-Newtonian
  time-domain waveforms.  Note in particular the discrepancy between
  T1 and T4, as these were used in the majority of the hybrid
  waveforms. On the left overlaps are calculated against the early
  aLIGO noise curve, on the right overlaps are calculated
  against the Zero-detuned, high-power noise curve.  All waveforms
  include terms up to 3.5 pN-order, as do the post-Newtonian portions
  of the hybrid waveforms.
 }
\end{figure}

Before discussing overlaps between the submitted NINJA hybrid
waveforms, we first present comparisons between time-domain
post-Newtonian waveforms of the kind used to construct hybrid
waveforms. (See the Appendix for a summary of the pN approximants used.)
These waveforms terminate at $M\omega$=0.136, 0.114,
0.069, and 0.135 for T1, T2, T3 and T4 respectively.  At $M=10
M_\odot$ these correspond to termination frequencies of \unit[435]{Hz},
\unit[369]{Hz}, \unit[222]{Hz} and \unit[439]{Hz}, respectively.  The results of our overlap calculations
are shown in Fig.~\ref{f:pn_overlaps}.  Ref.~\cite{Buonanno:2009zt} 
contains a far more extensive comparison between post-Newtonian approximants, 
although that analysis differs in several important aspects from what has 
been done here, notably the overlaps are maximized over the intrinsic mass
parameters of one waveform.  However the qualitative conclusion that
T1 matches better with T2 and T4 than with T3 is consistent with our
results.  These results should be borne in mind when evaluating
overlaps between hybrid waveforms using different pN approximants.  In
particular, overlaps at lower masses will be dominated by the influence of
the two pN waverforms, and therefore hybrid waveforms using T3 will have
relatively low matches against hybrid waveforms using T1.  Note also
that T3 and T4 diverge from T1 as mass increases, this is precisely
why we need to transition to numerical-relativity waveforms.

\begin{figure}
 \includegraphics[width=0.5\linewidth]{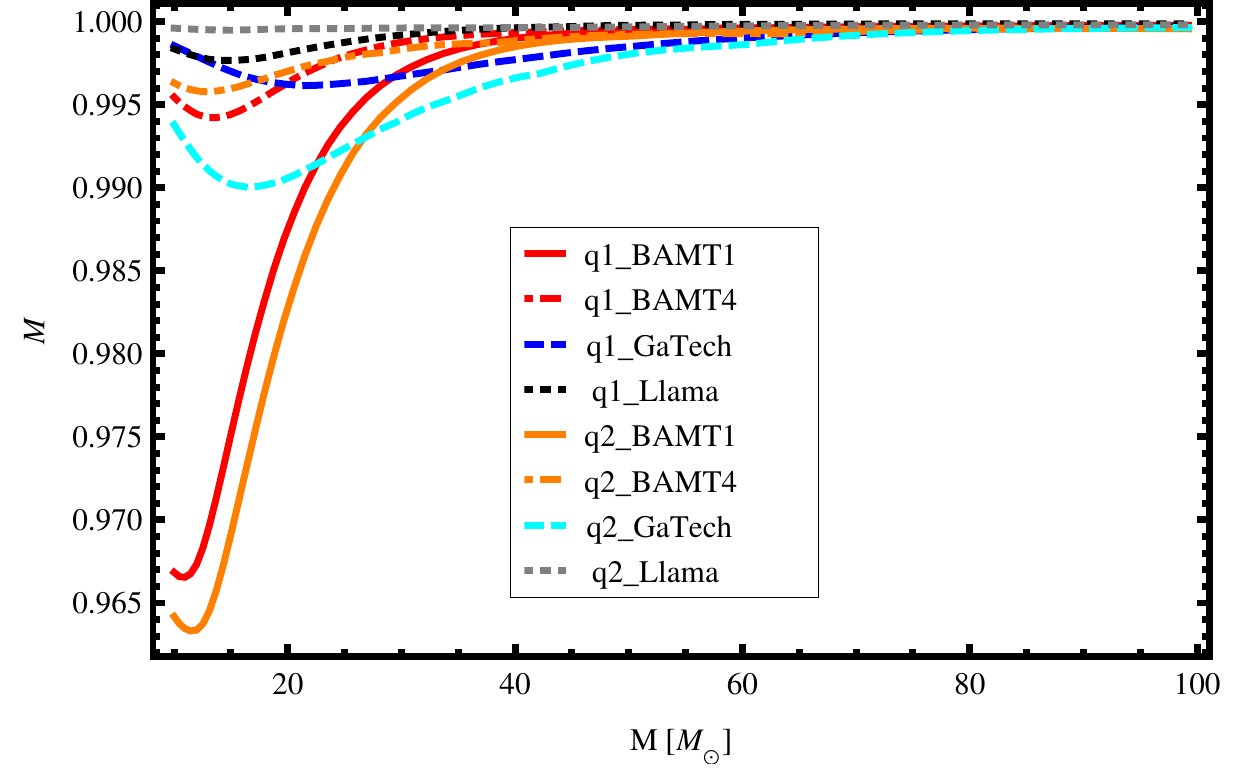}
 \hspace*{-0.03\linewidth}
  \includegraphics[width=0.5\linewidth]{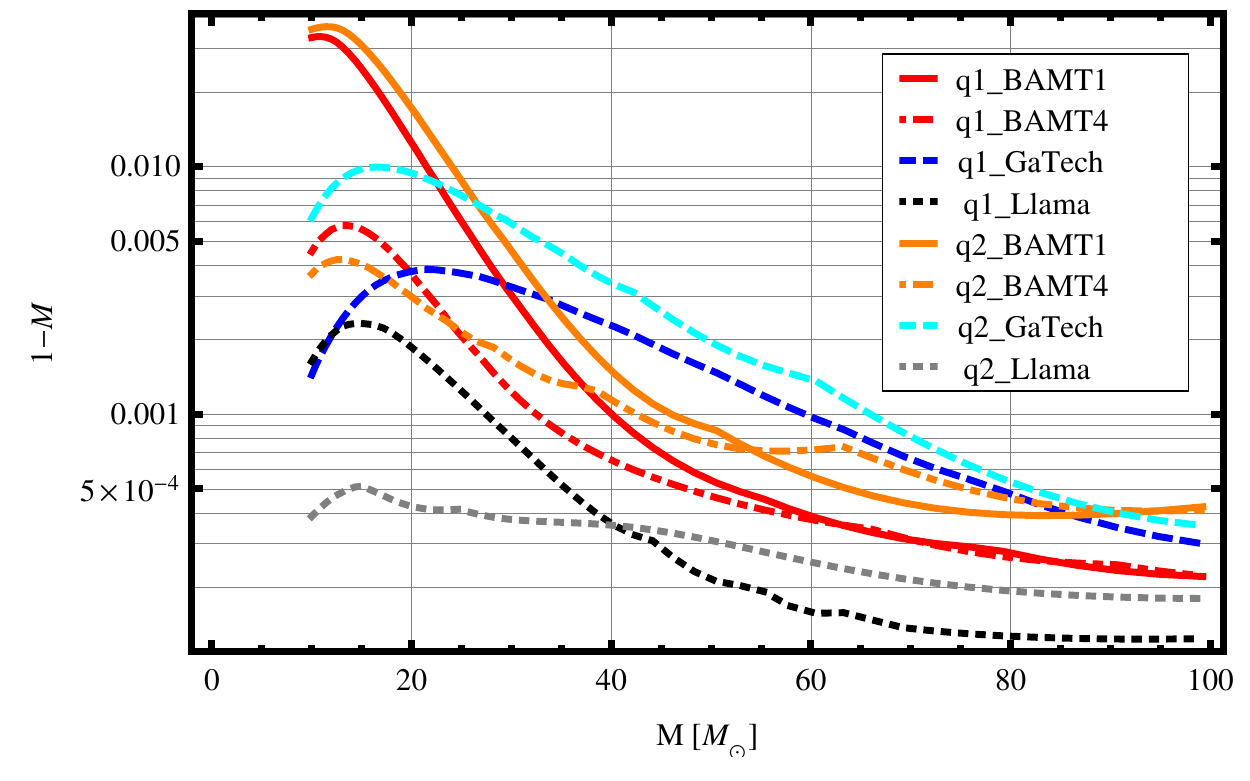}\\
  \includegraphics[width=0.5\linewidth]{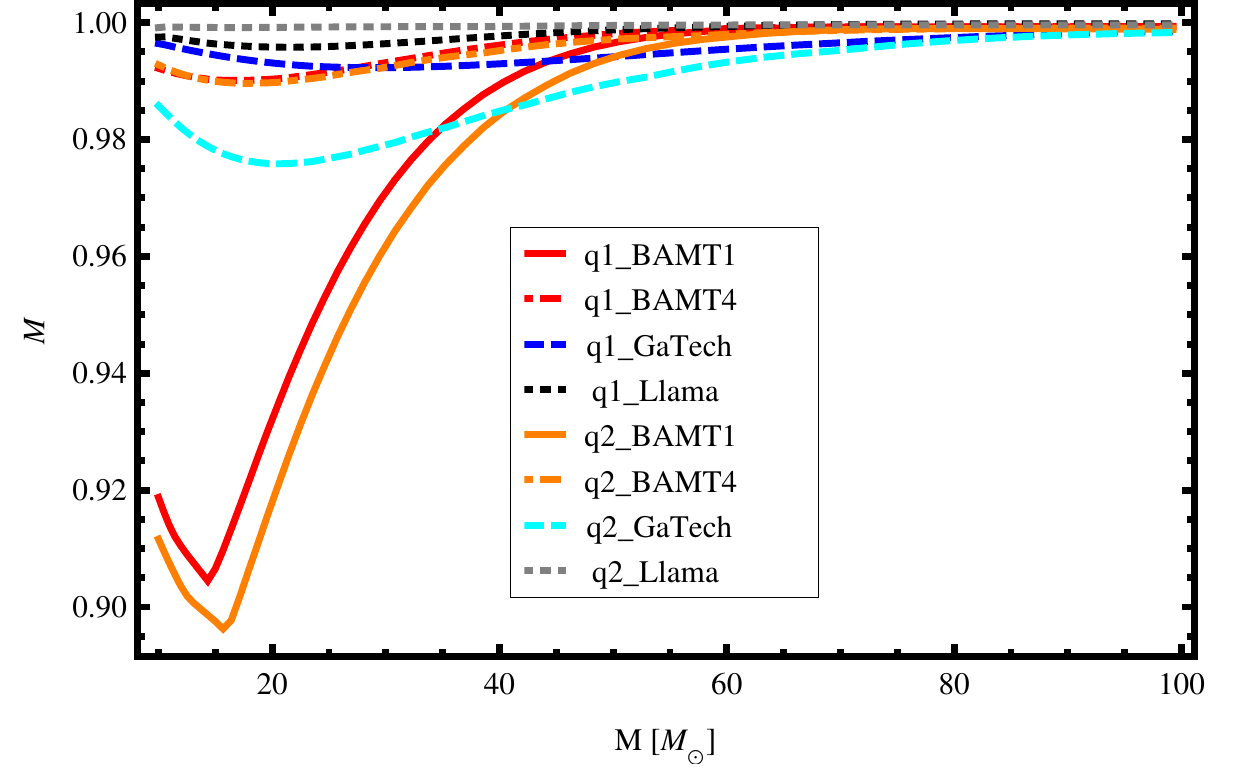}
 \hspace*{-0.03\linewidth}
  \includegraphics[width=0.5\linewidth]{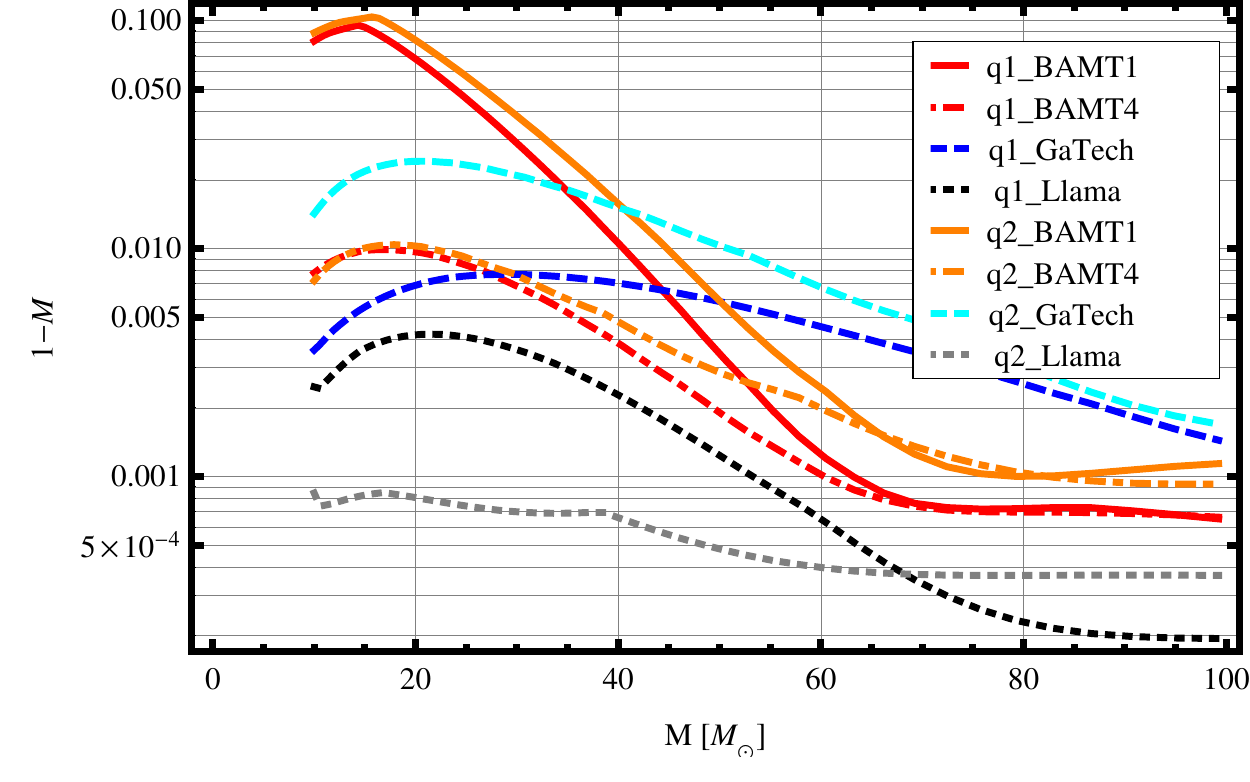}
  \caption[Overlaps between NINJA-2 submissions maximized over time and phase]{
  \label{f:ninja2_overlap_test}
  {\bf Differences between hybrid-waveforms for the same configuration}.  Plotted are the overlaps of the {\tt SpEC} (TaylorT4) hybrid against the
  four other submissions: All plots show results for both equal mass and mass ratio two
  hybrid waveforms with zero spin.  On the top
  row overlaps were computed using the early aLIGO noise curve.  On
  the bottom row the Zero-Detuned, High-Power noise curve was used.
  The overlaps are above 0.98 for identical pN--approximants.  For 
  different pN--approximants overlaps are above 0.98 for the early
  noise curve and above 0.90 for the Zero-Detuned, High-Power noise
  curve.
}
\end{figure}

\begin{figure}
 \includegraphics[width=0.5\linewidth]{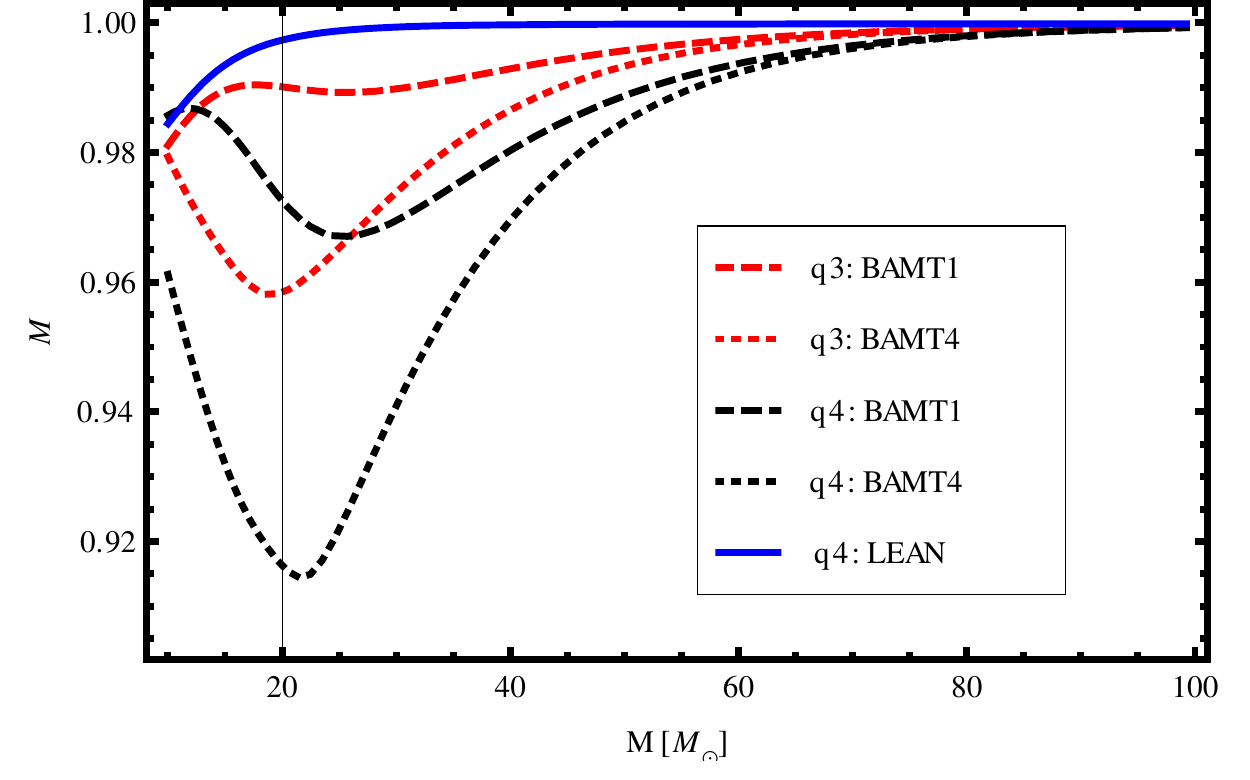}
 \hspace*{-0.03\linewidth}
  \includegraphics[width=0.5\linewidth]{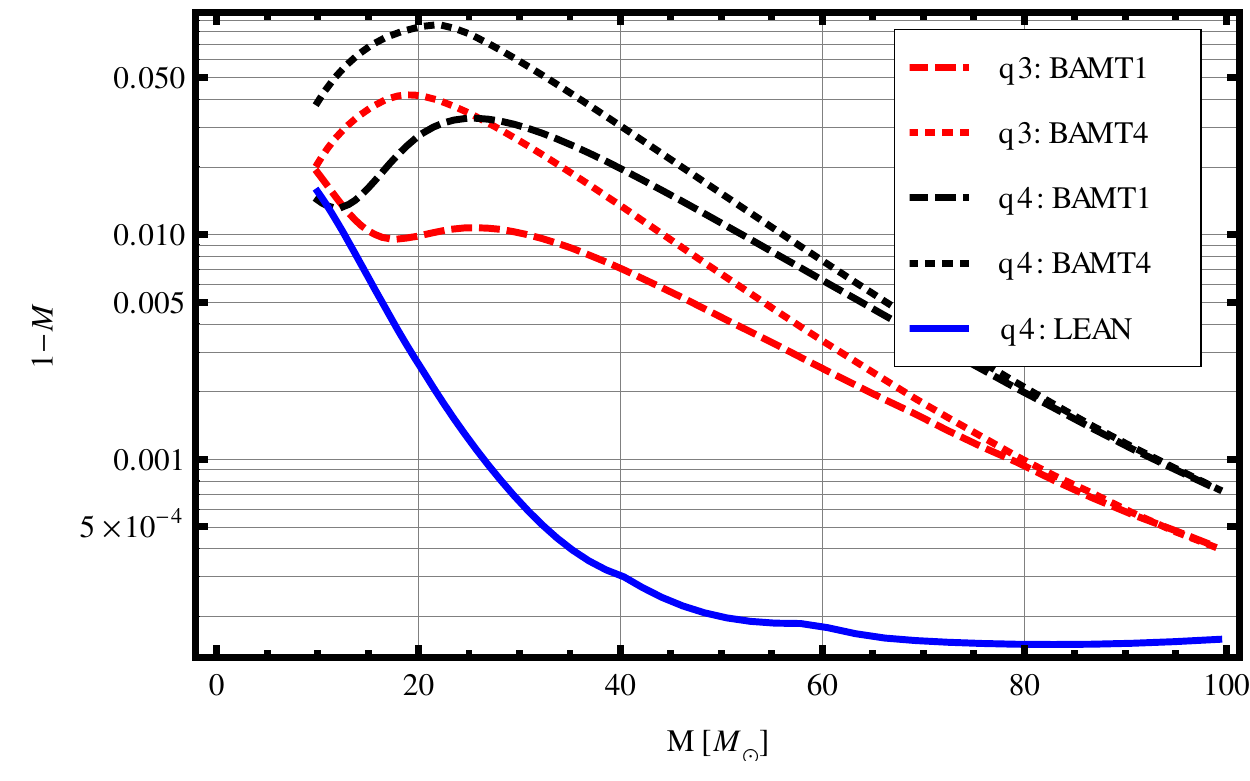}\\
  \includegraphics[width=0.5\linewidth]{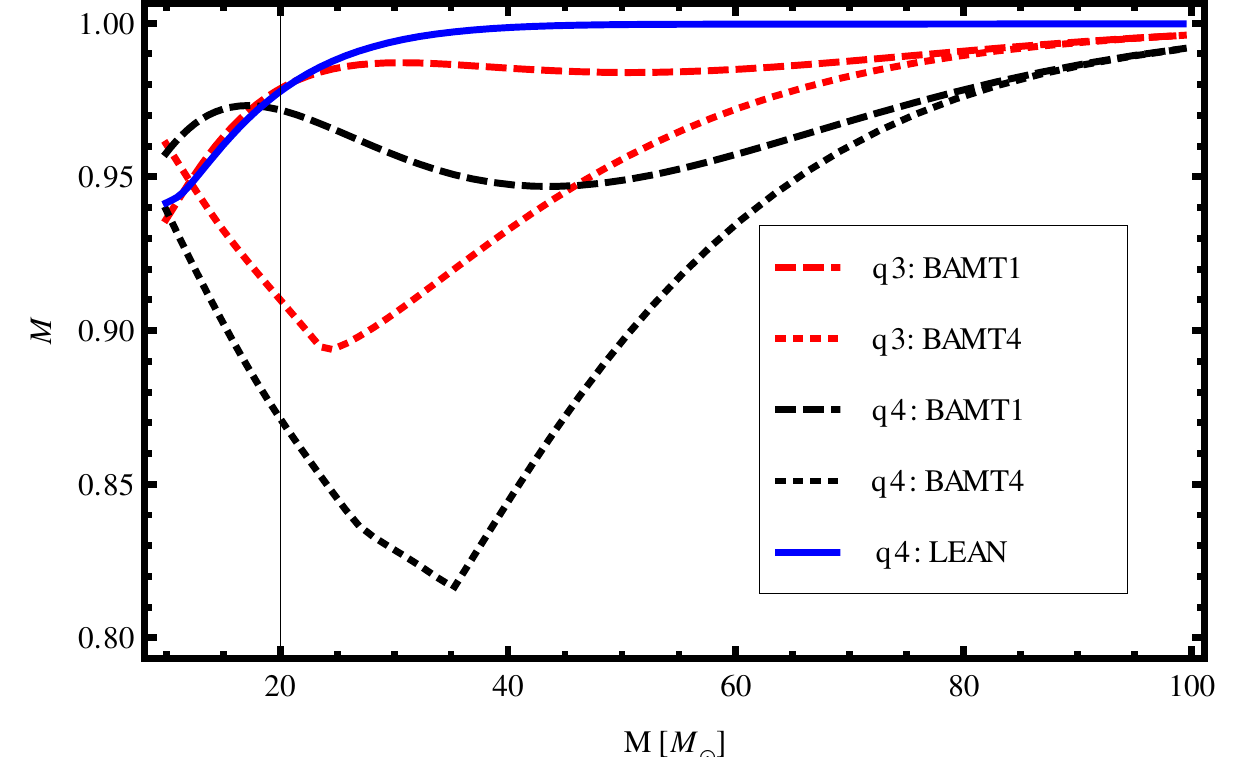}
 \hspace*{-0.03\linewidth}
  \includegraphics[width=0.5\linewidth]{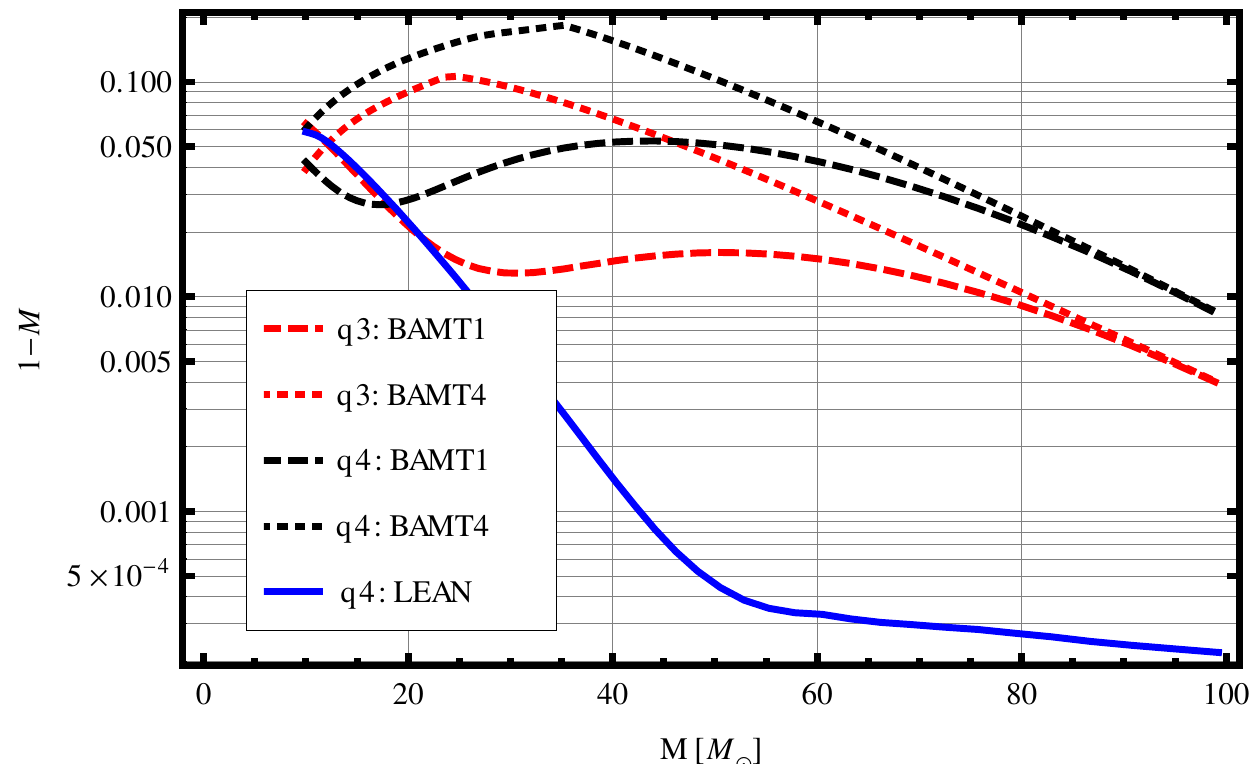}
  \caption[Overlaps between NINJA-2 submissions maximized over time and phase]{
  \label{f:ninja2_overlap_test_q34}
  {\bf Differences between hybrid-waveforms for the same configuration}.  Plotted are the overlaps of the {\tt SpEC} (TaylorT2) hybrid against the
  four other submissions: All plots show results for both mass ratio
three and four nonspinning
  hybrid waveforms.  On the top
  row overlaps were computed using the early aLIGO noise curve.  On
  the bottom row the Zero-Detuned, High-Power noise curve was used.
}
\end{figure}

\begin{figure}
 \includegraphics[width=0.5\linewidth]{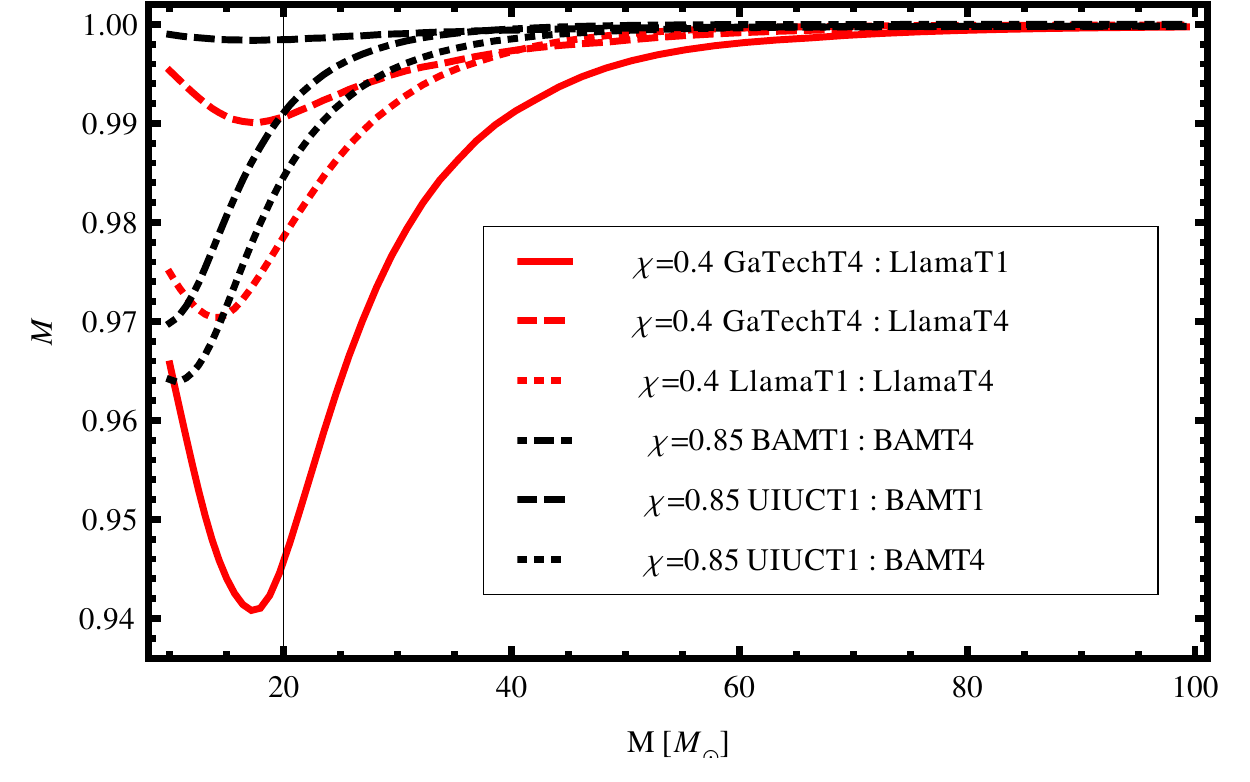}
 \hspace*{-0.03\linewidth}
  \includegraphics[width=0.5\linewidth]{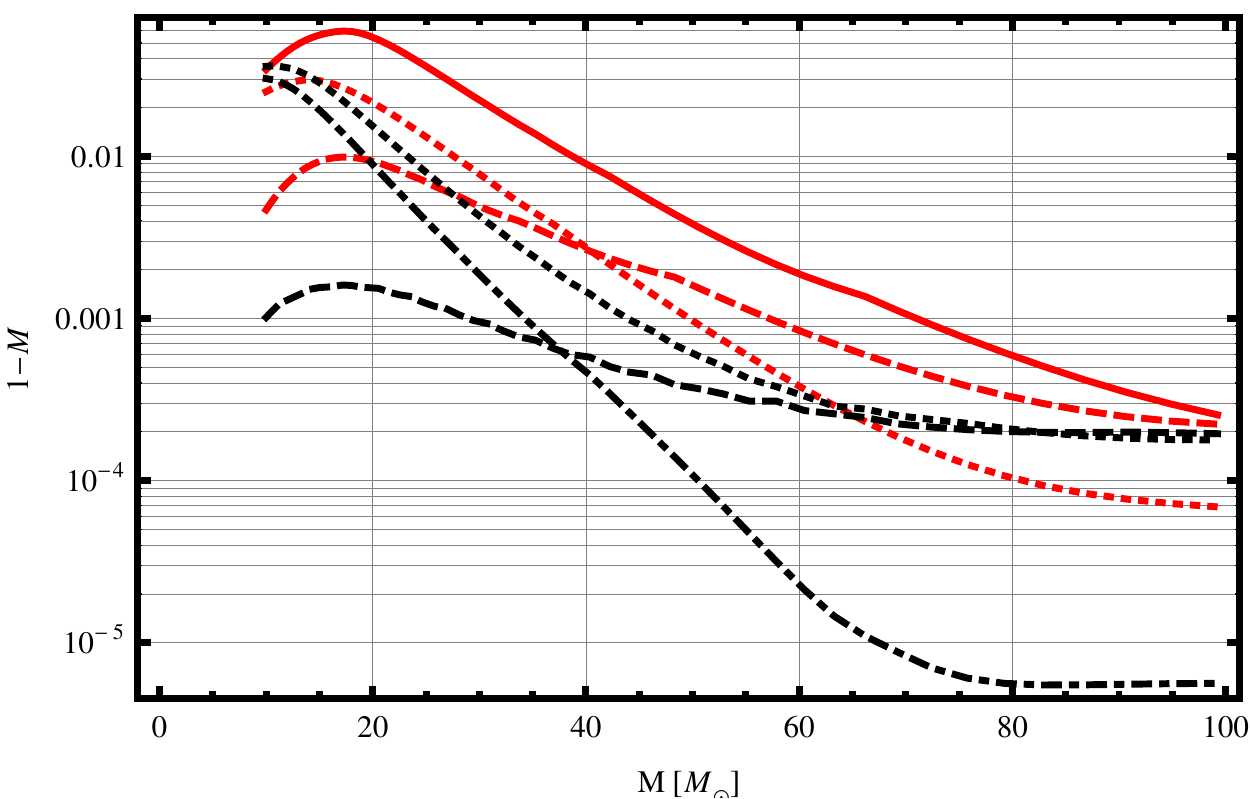}\\
  \includegraphics[width=0.5\linewidth]{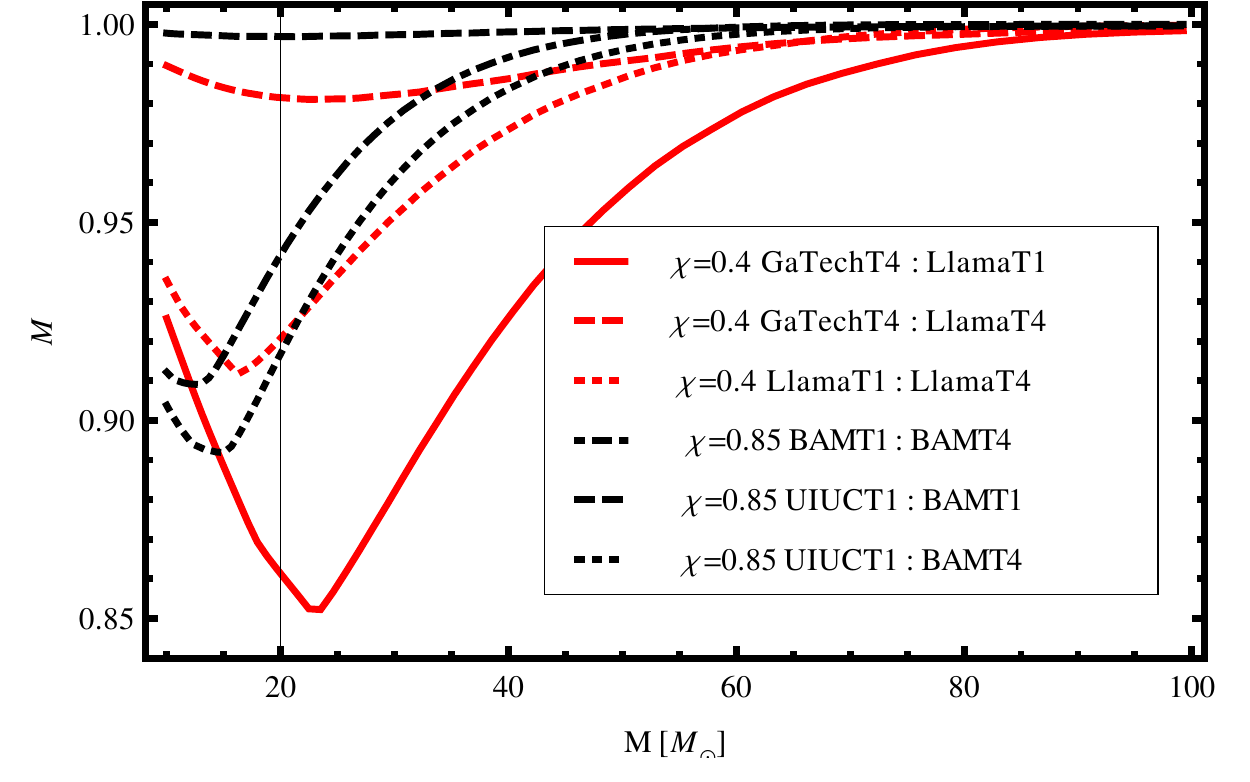}
 \hspace*{-0.03\linewidth}
  \includegraphics[width=0.5\linewidth]{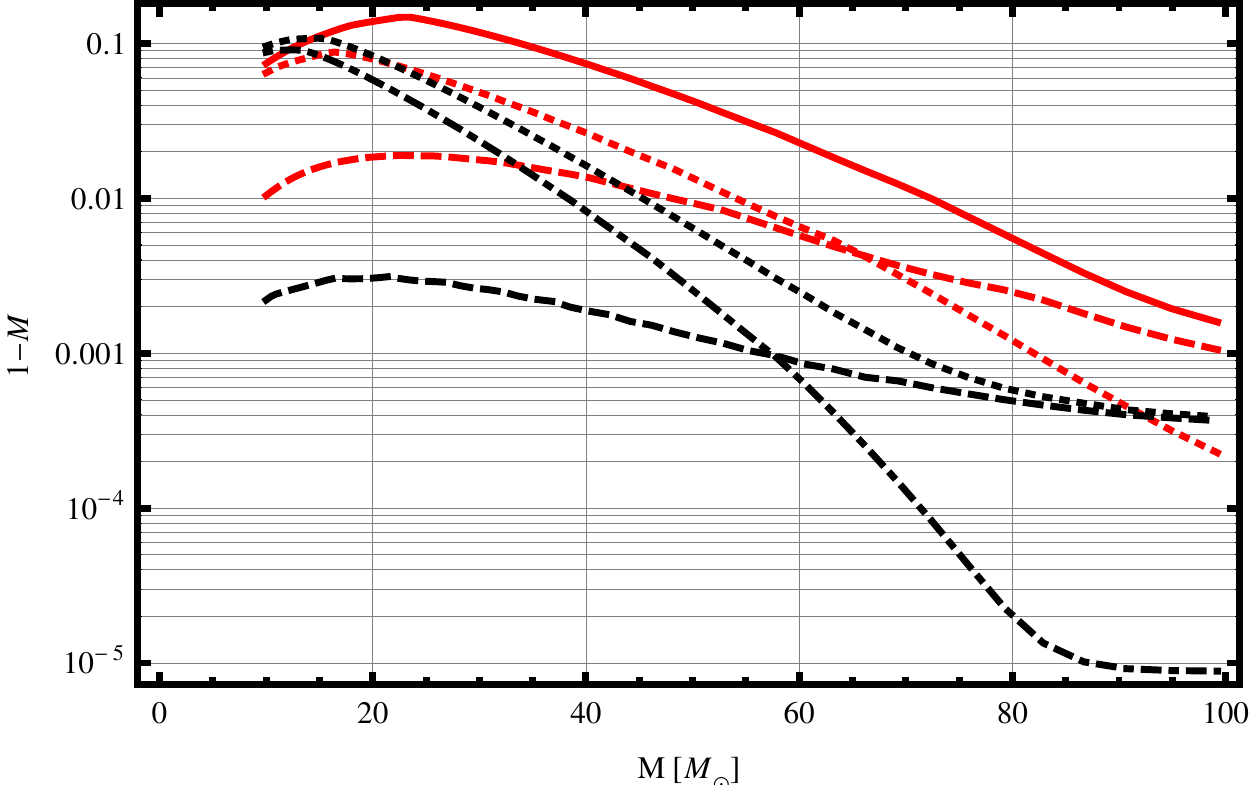}
  \caption[Overlaps between NINJA-2 submissions maximized over time and phase]{
  \label{f:ninja2_overlap_test_q1S}
  {\bf Differences between hybrid-waveforms for the same configuration}.  Plotted are the overlap comparisons of sets of equal mass spinning hybrids with $\chi_i = 0.4, 0.85$.
  On the top
  row overlaps were computed using the early aLIGO noise curve. On
  the bottom row the Zero-Detuned, High-Power noise curve was used.
  The overlaps are above 0.98 for identical pN--approximants.  For
  different pN--approximants overlaps deteriorate to 0.94 for the early
  noise curve and above 0.85 for the Zero-Detuned, High-Power noise
  curve.
}
\end{figure}

Let us now discuss the main results of this section, the overlap
between different submitted hybrid waveforms. For six of the black hole configurations
listed in Table~\ref{tab:ninja2_submissions}, hybrids have been
constructed independently for different  numerical waveforms,
and overlaps were computed between each pair
of waveforms in a group. Below we are showing overlap plots for
all six cases:  
The $q = \{1,2\}$ non-spinning waveform submissions are shown in
Fig.~\ref{f:ninja2_overlap_test},
    $q = \{3,4\}$ non-spinning waveforms in
Fig.~\ref{f:ninja2_overlap_test_q34},
and $q = 1$, $\chi_i = 0.4, 0.85$ spinning waveforms in
Fig.~\ref{f:ninja2_overlap_test_q1S}.
We will discuss the $q = \{1,2\}$ non-spinning cases in more detail.

At the high-mass end, where the NR portion of the waveform dominates
the overlap, the overlaps approach 1.  Since all these waveforms model
the same physical system, this is the expected behavior. High overlaps
at high masses indicate good agreement between different NR codes, and
the results here are consistent with the detailed study of the $q=1$ case was 
performed in the Samurai project~\cite{Hannam:2009hh}.  At
the low-mass end, where the overlap is dominated by the pN portion of
the waveform, the behavior is qualitatively as expected from
Fig.~\ref{f:pn_overlaps}.  In particular the overlap between the {\tt
BAM} hybrid using T1 and the {\tt SpEC} hybrid using T4 is lower than
the matches between the other hybrids, all of which use T4.  In the
region between $\roughly 15 - 30 \msun$ there are drops in the
matches between hybrids using T4.  In this mass range the
hybridization regions are passing through the most sensitive frequency
band, and the mismatches are due in part to different choices of
hybridization methods and parameters.  The question of how many NR
cycles are needed in order to produce a robust hybrid waveform is an
area of active
research~\cite{Hannam:2010ky,Damour:2010zb,MacDonald:2011ne,Boyle:2011dy,Ohme:2011zm}.

If the approximants are the same, then the mismatches
will depend only on the differences in (1) the hybridization methods,
(2) the hybridization frequencies (and windows), and (3) the NR data.
We have performed tests to verify that the overlap due to these effects is very small.
We have made overlap calculations using a white-noise PSD, integrated
between \unit[10]{Hz} and \unit[100]{Hz}, so the hybridisation region can pass fully into 
and out of band as the mass is varied between 10\,$M_\odot$ and 100\,$M_\odot$,
and found that the maximum mismatch due to hybridization is 0.05\% for
the GATech and SpEC equal-mass nonspinning hybrids . This suggests that the contribution of the
hybridization to the mismatch is very small, which is consistent with the 
results in~\cite{Hannam:2010ky}. Mismatches for masses $M \apprge 150\,M_\odot$ will
be due only to differences between the numerical data, and we find these to be
0.1\%, which is consistent with the results for equal-mass nonspinning 
binaries in~\cite{Hannam:2009hh} and for $q=2$ nonspinning binaries 
in~\cite{Santamaria:2010yb}.

The overlap plots discussed thus far do not yet address the accuracy required
for detection of gravitational waves.  Broadly, in order to claim a detection
a signal must match well against at least one template waveform used in a
search.  Several methods of assessing detection accuracy have been
proposed~\cite{Lindblom:2008cm,Lindblom:2010mh}, however here we take the
simple aproach that the waveforms are sufficient for further detection
studies if the overlap is above the the standard 0.97 threshold for a loss of
no more than 10\% of signals in a search.  In some cases above the overlaps
are not above this threshold, but the appropriate quantity to evaluate to
address this question is the overlap maximized over all of the physical
parameters in a search. 
We now extend these overlap studies by maximizing over one physical 
parameter, the mass of one of the waveforms, as well as the time and phase.  
If the overlaps are now all over 0.97 (which they are), then they are acceptable
for use in search-related studies. 

\begin{figure}
  \includegraphics[width=0.50\linewidth]{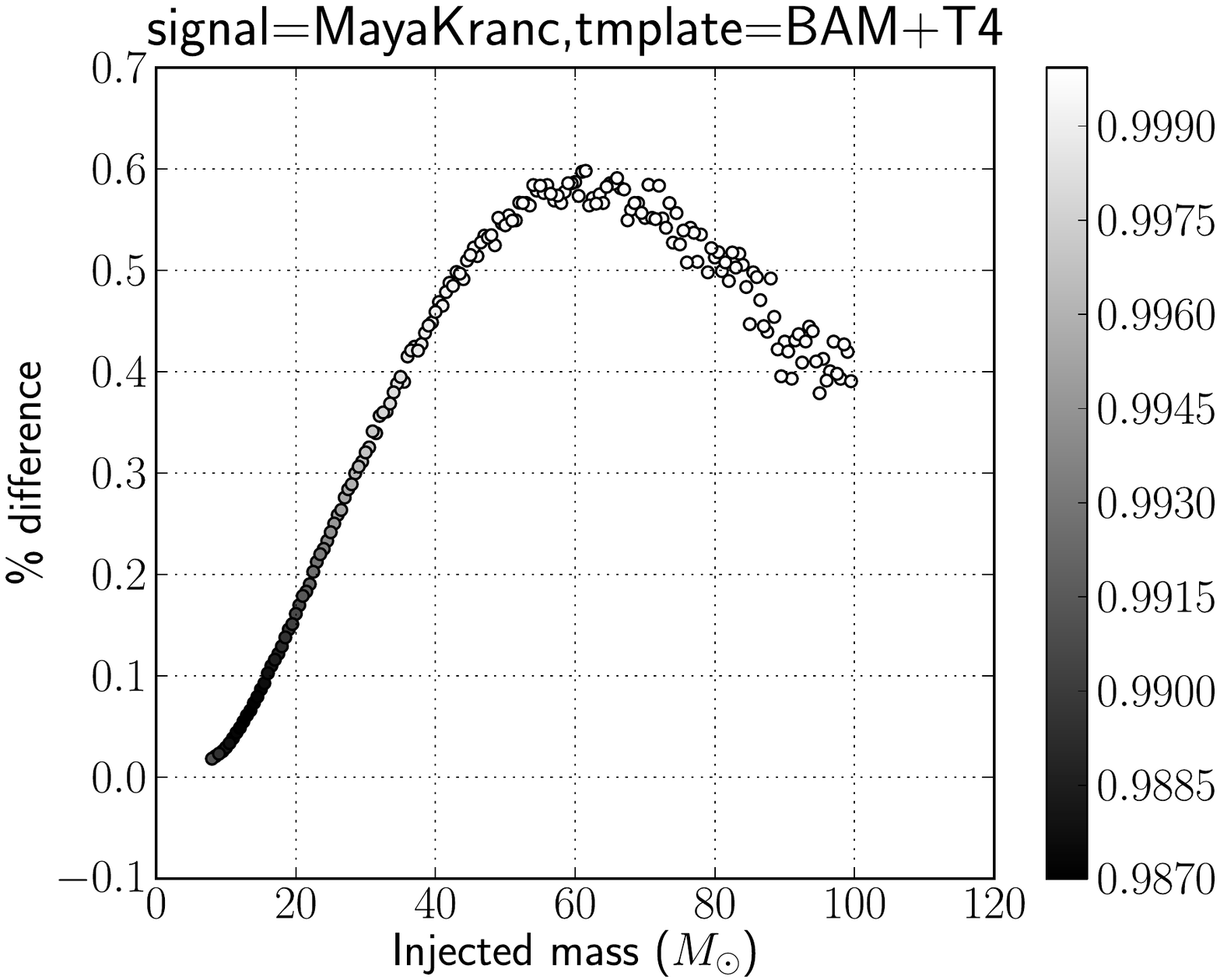}
  \hspace*{-0.05\linewidth}
  \includegraphics[width=0.50\linewidth]{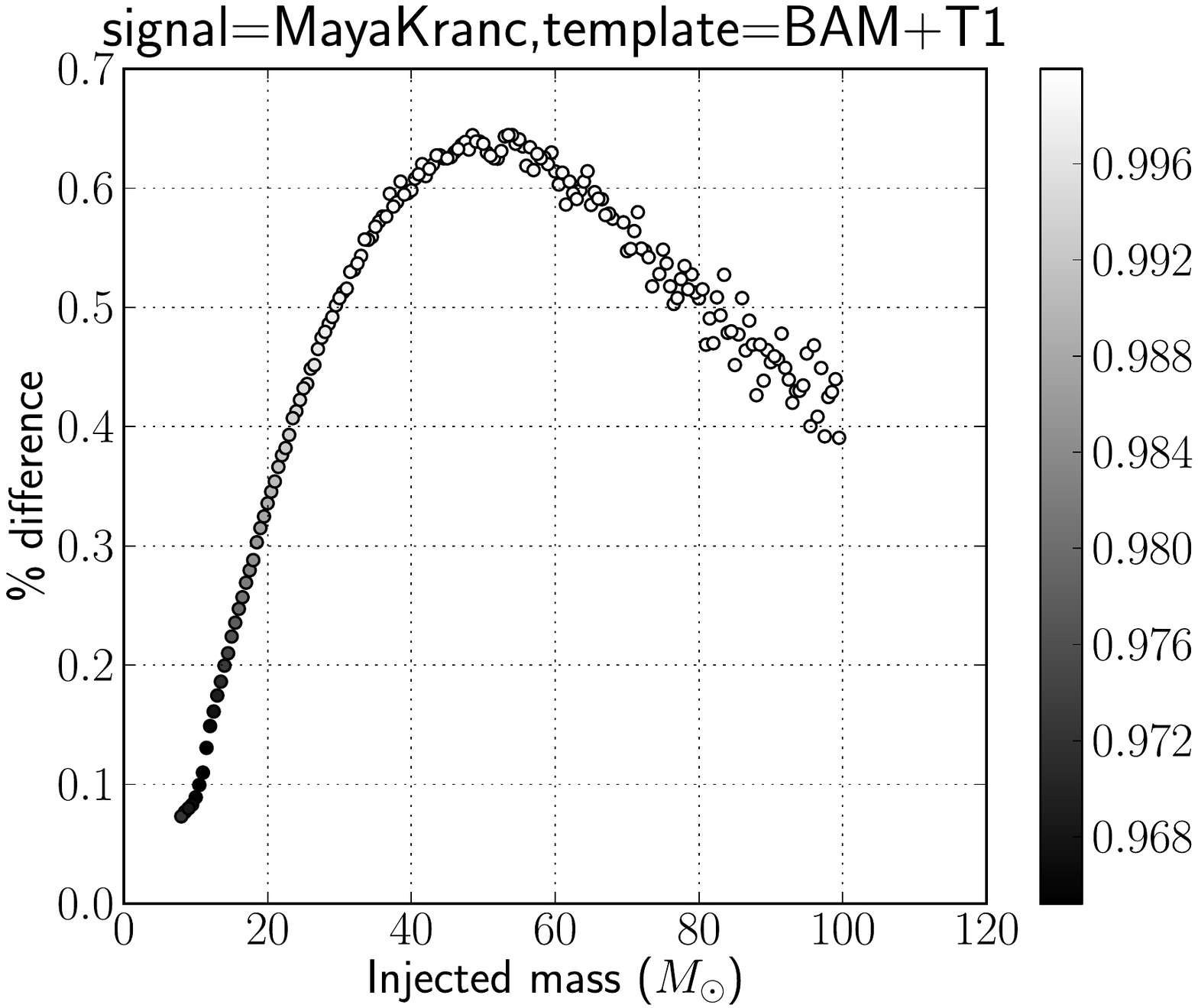}
  \caption[Overlaps between unequal-mass submissions maximized over mass]{
  \label{f:ninja2_max_over_mass_bam}
  Overlaps and mass-bias when searching for one hybrid waveform with
  another hybrid waveform of the same configuration.  {\bf Left panel:}
  the $q=1$ non-spinning {\tt MayaKranc} waveform is taken as the
  signal, and the $q=1$ non-spinning {\tt BAM} waveform hybridized with
  TaylorT4 taken as the template.  {\bf Right panel:}  the $q=1$,
  non-spinning {\tt MayaKranc} waveform is taken as the signal and the
  $q=1$ non-spinning {\tt BAM} waveform hybridized with TaylorT4 is
  taken as the template.  In both panels, maximization is done over the
  mass of the template, as well as over time and phase.  The horizontal
  axis gives the mass of the signal; the vertical axis gives the
  fractional difference between the injected mass and the mass of the
  template that maximizes the overlap.  Overlaps are calculated against
  the early aLIGO noise curve.
 }
\end{figure}

These extended overlaps also provide insight into the 
``parameter estimation question,'' as the error in parameter
recovery between two hybrid waveforms gives a rough lower bound on the
errors we may expect in recovering parameters of hybrid injections
with search templates. 
In practice parameter recovery 
is likely to be worse than this, as it will involve maximizing 
over several parameters in addition to total mass, in addition to differences
between the waveforms.  Example plots
using the equal-mass, non-spinning {\tt MayaKranc} waveform as the
signal and {\tt BAM} hybridized with two different pN approximants as
the template are shown in Fig.~\ref{f:ninja2_max_over_mass_bam}. Note that
in the right panel of Fig.~\ref{f:ninja2_max_over_mass_bam} the minimum overlap
is below 0.97. This minimum match will be even worse for $q>1$ 
binaries~\cite{Hannam:2010ky}, but if maximization is done over the
other physical parameters (mass ratio and spin), then the overlap will increase to
well above 0.97~\cite{Ohme:2011zm}.

\begin{figure}
  \includegraphics[width=0.50\linewidth]{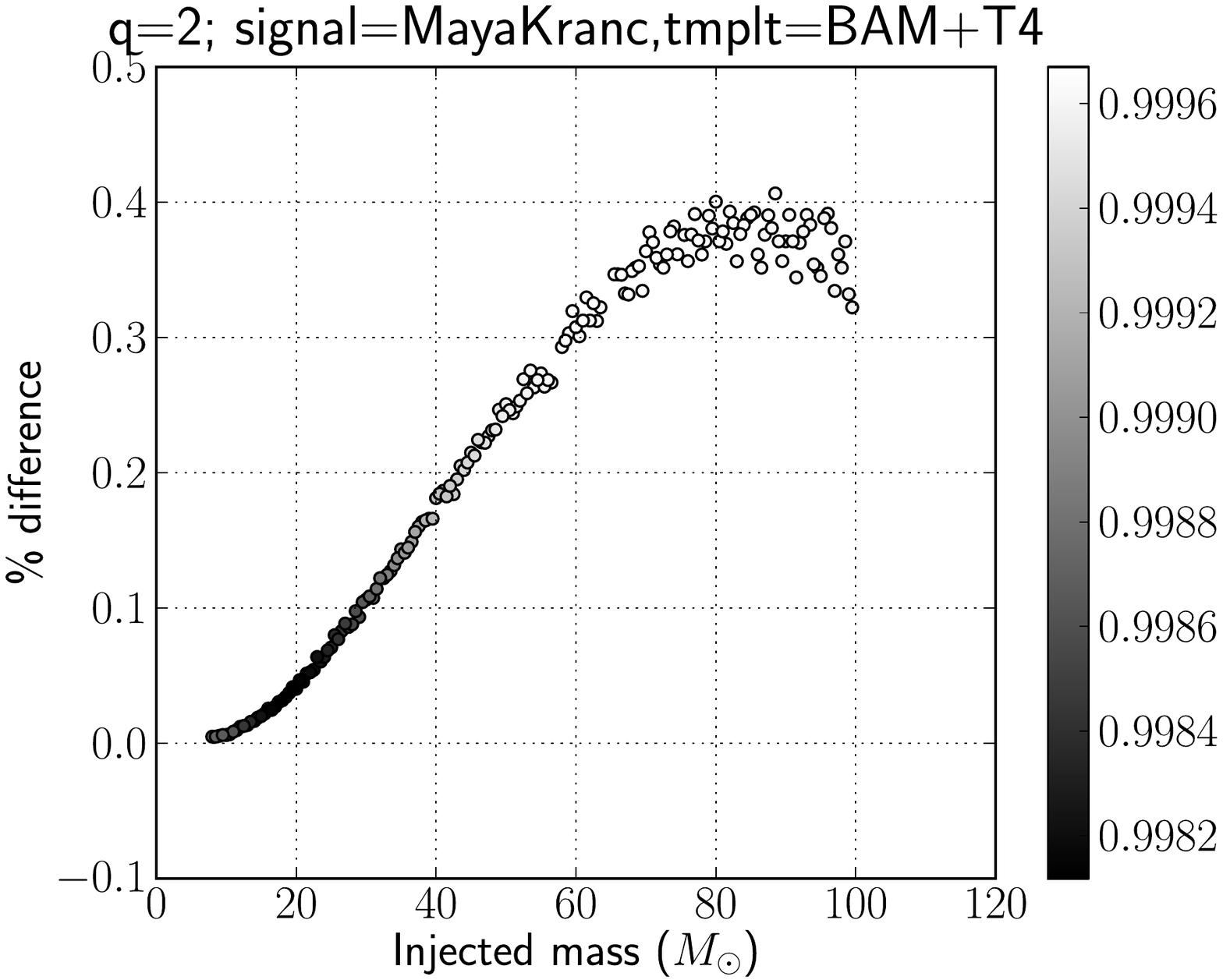}
  \hspace*{-0.05\linewidth}
  \includegraphics[width=0.50\linewidth]{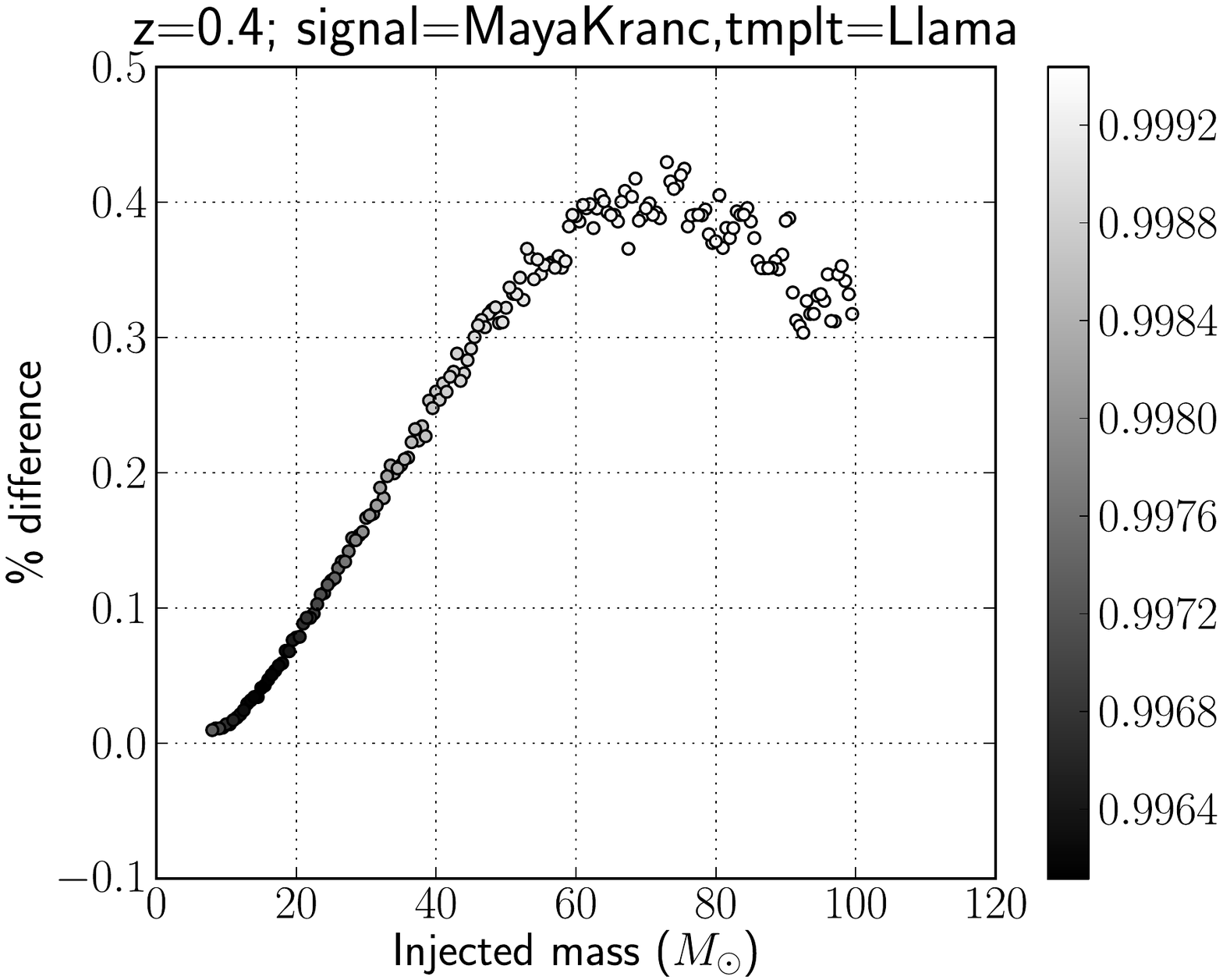}
  \caption[Overlaps between unequal-mass submissions maximized over mass]{
  \label{f:max_over_m_q2}
Overlaps and mass-bias when searching for one hybrid waveform with
another hybrid waveform of the same configuration.  {\bf Left panel:}
the $q=2$ non-spinning {\tt MayaKranc} TaylorT4 waveform is taken as
the signal, and the $q=2$ non-spinning {\tt BAM} waveform hybridized
with TaylorT4 taken as the template.  {\bf Right panel:}  the $q=1$,
$\chi_1=\chi_2=0.4$ {\tt MayaKranc} Taylor T4 waveform is taken as signal and
searched for with the Llama waveform of the same parameters and pN
approximant.  In both panels, maximization is done over the mass of
the template, as well as over time and phase.  The horizontal axis
gives the mass of the signal; the vertical axis gives the fractional
difference between the injected mass and the mass of the template that
maximizes the overlap.  Overlaps are calculated against the early
aLIGO noise curve. }
\end{figure}

At the high-mass end the overlap is dominated by NR data, and as in
Fig.~\ref{f:ninja2_overlap_test} the overlaps are high without needing
to move off the signal mass.  At the low-mass end the same result
would be expected in a pure pN/pN comparison although there is enough
of the hybridization in-band to reduce the overlaps.  However,
changing the mass introduces a phase difference that accumulates over
all the cycles in-band, and so higher overlaps cannot be achieved.
The result is optimal mass values close to the correct mass value, but
with a lower overlap.  In the middle region these factors compete.  At
higher masses the overlap is reduced less by changing the mass and so
the recovered value of the mass can stray further from the injected
value.  As the hybridization passes out of band this adjustment is no
longer needed.  The same general behavior can be seen in comparisons
between non-spinning, unequal-mass ($q=2$) waveforms, and for
equal-mass spinning waveforms, as shown in Fig.~\ref{f:max_over_m_q2}
which shows overlaps between waveforms with identical parameters and
hybridized with the same pN approximants maximized over the mass of
one waveform.  We can infer from these results that, although we have
only made a crude estimate of the parameter bias, the accuracy of the
waveforms (excluding the uncertainty in the pN approximants) is
extremely high.  There are many factors that may bias parameter
estimation of real signals, including uncertainties in the detector
calibration, noise in the detectors and errors in pN waveforms used as
templates.  Subsequent NINJA-2 data analysis studies will attempt to
quantify the effects of these factors.

\section{Conclusion}
\label{sec:discussion}
The efficiency of searches for GW signals from BH binaries and the
accuracy with which the source parameters are estimated will depend crucially on
progress in incorporating information from approximation methods and numerical relativity into
the waveform models used in data-analysis algorithms. Even in the nonspinning case, a 
recent extensive comparison of different pN approximants~\cite{Buonanno:2009zt} concludes 
that for masses above 12\,$M_\odot$ numerical-relativity simulations of the last orbits and
merger are required for the construction of optimal detection templates. This need is expected 
to be even greater when spinning binaries are included.  And of course, numerical-relativity simulations will be yet more important for parameter estimation. 

Injections of inspiral-merger-ringdown (IMR) waveforms from analytical waveform models are 
already used to calibrate the analysis of detector data~\cite{Abadie:2011kd}. However, the synthesis of
NR results into waveform models typically lags several years behind the
most complete current sets of NR waveforms. The NINJA-2 project will consider direct 
injections of hybrid pN-NR waveforms, which will allow the use of waveforms which have not 
(yet) been used in constructing analytical models, and will avoid any additional modeling errors. 
Using the best available IMR waveforms will be particularly important for parameter estimation.

The first goal of the second NINJA project has been to review the submitted hybrid waveforms. 
In this paper we have summarized the requirements that the submitted waveforms must meet
(see Sec.~\ref{sec:introduction}). All submitted waveforms meet these requirements, within some 
minor caveats that are detailed in Sec.~\ref{sec:introduction}. We have
also demonstrated the validity of the hybrid waveforms for use in the
NINJA-2 project with more detailed consistency checks of the submitted data.

The validation of and comparisons between the submissions, which we have presented here,
are a major feature of NINJA-2 that was not present in NINJA-1, and will be indispensable 
when analyzing the results from systematic injection studies over the course of the NINJA-2 project.
A total of 63 waveforms from 8 different groups were contributed to NINJA-2, corresponding to 46 
distinct NR waveforms, and 29 different configurations of mass ratio and spins.  For six configurations, 
multiple numerical waveforms were submitted, and 16 numerical waveforms were hybridized with two 
different pN approximants (TaylorT1 and TaylorT4). This has allowed waveform comparisons and tests 
of the accuracy standards, as discussed in detail in Sec.~\ref{sec:comparisons}, and summarized below.
 For each submission, significant further information has been included in metadata files as specified 
 in~\cite{Brown:2007jx}, e.g., hybridization frequencies, eccentricities, or literature references. This 
 allowed us to automatically generate information such as Table~\ref{tab:ninja2_submissions} and 
 has proved crucial in systematically analyzing the data set. 
Verifying these submissions, preparing a consistent data set, and evaluating their accuracy by 
comparing submissions with the same mass-ratio and spin, has been a formidable task, and in
the present paper we {\em only} report on the $(\ell, m)=(2,2)$ spherical-harmonic modes.   Based on 
our various checks and comparisons as reported in this paper, we judge the submitted waveforms 
suitable for the NINJA-2 project.  
Parameter estimation places particular stringent demands on waveform templates.  We believe the 
NINJA-2 waveforms to be of sufficient quality to estimate size and shape of maximum likelihood 
contours.  However, the large truncation error of the post-Newtonian expansion in the cycles before 
and during hybridization will induce systematic errors; further investigations will be necessary before 
these waveforms can be used to check the accuracy of the estimated parameters themselfes (i.e. the 
location of the maximum likelihood contour, rather than its shape).

As seen in Fig.~\ref{f:ninja2_param_map}, the submissions 
primarily cover only two lines  in the ($q$, $\chi$) plane,
leaving large regions of parameter space unexplored.  We also do not
consider any precessing signals, which adds several dimensions to the
parameter space. These issues will be explored in future NINJA projects. Extending our analysis to 
non-dominant spherical-harmonic modes, and to a larger volume of parameter space, 
will mark a significant challenge for the future, and will require 
a significant extension of our methods of analysis, and of automatizing them, and will constitute a substantial research project by itself. 

An important shortcoming in our analysis of waveform accuracy is that
it is currently not possible to accurately estimate the pN truncation error. A reasonable approach 
seems to be to compare pN approximants, which are accurate to the highest known order, but 
differ in terms beyond this order (3.5pN without spins). These deviations still depend strongly on 
the choice of pN approximants used in any comparison, and on the mass ratio and spins.
For example, in Figs.~\ref{f:pn_overlaps}, \ref{f:ninja2_overlap_test} and \ref{f:ninja2_max_over_mass_bam}
we compare the use of the TaylorT1 and TaylorT4 approximants 
for mass ratio $q=1$ and zero spins, Fig.~\ref{f:ninja2_overlap_test}
also includes $q \neq 1$. Previous work implies that TaylorT4 performs exceptionally well in these cases~\cite{Boyle:2007ft,Hannam:2010ec}, apparently by 
coincidence, but beyond that it is not possible to make precise quantitative statements; this 
figure simply illustrates the general level of mismatch error that we may expect between 
various pN approximants across a range of binary masses.

We caution that the comparisons presented in
Sec.~\ref{sec:comparisons} were done only for subset of the black hole configurations, which
do not include the most extreme cases, and not the cases with unequal Kerr parameters.
Further work will be necessary to establish
with similar confidence that all NINJA-2 hybrid waveforms 
are of similar quality.

Our waveform comparisons in Sec.~\ref{sec:comparisons} are consistent with previous results: mismatches between 
waveforms are dominated by the choice of pN approximant, while NR and hybridization errors are far smaller. 
Hybridization choices and methods certainly affect overlaps, although the mismatch due to hybridization appears to be less than a fraction of a percent. This is not expected to have any noticeable effect on detection, but is likely to impact parameter estimation.  The degree to which these effects will bias searches is one of the key question we hope NINJA-2 will be able to answer.
In particular, theoretical studies such as the ones presented in this paper cannot replace a 
complete parameter estimation analysis, and it is not clear how theoretical studies based on 
Gaussian noise can predict the performance of GW searches in real noise.
 Another important goal of  the NINJA-2 project is thus to build up experience in comparing simple theoretical studies with the full GW-search plus parameter estimation pipeline exercised in actual observations of GW events.

\section*{Acknowledgments}

We thank Ilya Mandel for helpful discussions.

%
%
We acknowledge gratefully support from the National Science Foundation
under NSF grants
PHY-1040231, 
PHY-0600953, 
PHY-0847611, 
AST-1028087, 
DRL-1136221, 
OCI-0832606, 
PHY-0903782, 
PHY-0929114, 
PHY-0969855, 
AST-1002667, 
PHY-0650377, 
PHY-0963136, 
PHY-0855315, 
PHY-0969111, 
PHY-1005426, 
PHY-0601459, 
PHY-1068881, 
PHY-1005655, 
PHY-0653550, 
PHY-0955773, 
PHY-0653443, 
PHY-0855892, 
PHY-0914553, 
PHY-0941417, 
PHY-0903973, 
PHY-0955825, 
NSF cooperative agreement PHY-0757058, 
%
by NASA grants 
07-ATFP07-0158, 
NNX07AG96G, 
NNX10AI73G, 
NNX09AF96G, 
NNX09AF97G, 
%
%
%
by Marie Curie Grants of the 7th European Community Framework Programme
FP7-PEOPLE-2011-CIG CBHEO No.~293412, by the DyBHo-256667 ERC Starting Grant, 
and
MIRG-CT-2007-205005/PHY, 
and Science and Technology Facilities Council grants ST/H008438/1
and ST/I001085/1.
%
%
Further funding was provided by
the Sherman Fairchild Foundation, 
NSERC of Canada,  
the Canada Research Chairs Program, 
the Canadian Institute for Advanced Research, 
Govern de les Illes Balears, 
the Ram{\'o}n y Cajal Programme of the Ministry of Education and Science of Spain, 
contracts AYA2010-15709, CSD2007-00042, CSD2009-00064 and FPA2010-16495 of the Spanish Ministry of Science and Innovation, 
the Royal Society, 
and the Research Corporation for Science Advancement. 
%
%
Computations were carried out on Teragrid machines Lonestar, Ranger,
Trestles and Kraken under Teragrid allocations 
TG-PHY060027N, 
TG-MCA99S008, 
TG-PHY090095, 
TG-PHY100051, 
TG-PHY990007N, 
TG-PHY090003,  
TG-MCA08X009. 
Computations were also performed on the clusters 
```HLRB-2'' at LRZ Munich,
``NewHorizons'' at RIT (funded by NSF Grant Nos.~AST-1028087, DMS-0820923 and PHY-0722703), 
``Zwicky'' at Caltech (funded by NSF MRI award PHY-0960291),
``Finis Terrae'' (funded by CESGA-ICTS-2010-200),
``Caesaraugusta'' (funded by BSC Grant Nos.~AECT-2011-2-0006, AECT-2011-3-0007),
``MareNostrum''  (funded by BSC Grant Nos.~AECT-2009-2-0017, AECT-2010-1-0008, AECT-2010-2-0013, AECT-2010-3-0010, AECT-2011-1-0015, AECT-2011-2-0012),
``VSC'' in Vienna (funded by FWF grant P22498),
``Force'' at GaTech, 
and on the GPC supercomputer at the SciNet HPC Consortium; SciNet is
funded by: the Canada Foundation for Innovation under the auspices of
Compute Canada; the Government of Ontario; Ontario Research Fund -
Research Excellence; and the University of Toronto.

\appendix

\section{Post-Newtonian Waveforms}
\label{sec:pn_waveforms}

The accuracy of hybrid waveforms depends very sensitively on the
accuracy of the post-Newtonian waveforms with which they are
constructed.  For this reason, the NINJA-2 collaboration invested
significant effort into ensuring that all contributions used the most
current pN information available.  Mathematica notebooks containing
the full expressions and derivations of the various approximants are
provided in the ancillary files available with this paper online.
Here, we describe the techniques.

The pN approximants used in this paper solve for the orbital motion to
high accuracy by assuming that the motion is an adiabatic
quasicircular inspiral, and by assuming that the loss of orbital
binding energy $E$ during inspiral is balanced by the flux of energy
in gravitational radiation to infinity $\mathcal{F}$ and the flux of
energy going into the individual black holes caused by tidal heating
$\dot{M}$.  The first assumption means that the motion can be
described completely by the orbital phase function $\Phi(t)$.  We then
define the pN expansion parameter $v \define (M\, \rmd\Phi/\rmd
t)^{1/3}$, where $M$ is the sum of the apparent-horizon masses of the
black holes.  The orbital binding energy, gravitational-wave flux, and
tidal heating can be expressed as functions of this parameter.  Thus,
the energy-balance equation can be written as $\dot{E} + \mathcal{F} +
\dot{M} = 0$.  Using the chain rule to rewrite $\dot{E}$, we can
rearrange this as an expression for $\rmd v / \rmd t$, and include the
expression for $\rmd \Phi / \rmd t$ to obtain a complete system of
ordinary differential equations describing the motion of the binary:
\begin{equation}
  \label{eq:PNODEs1}
  \frac{\rmd v}{\rmd t} = - \frac{\mathcal{F}(v) + \dot{M}(v)} {E'(v)}
  \qquad \frac{\rmd \Phi} {\rmd t} = \frac{v^{3}}{M}.
\end{equation}
The formulation of the pN approximants, then, comes down to writing
down the expressions for $E(v)$, $\mathcal{F}(v)$, and $\dot{M}(v)$,
and integrating the system for $v(t)$ and $\Phi(t)$.

At leading order, the expressions for $E$ and $\mathcal{F}$ are
\begin{equation}
  \label{eq:OrbitalEnergy}
  E(v) = -\frac{M \eta\, v^2}{2} \qquad \text{and} \qquad
  \mathcal{F}(v) = \frac{32}{5}\, v^{10}\, \eta^2,
\end{equation}
where higher-order terms include additional factors of $v$.  The
additional terms are currently known up to $v^{7}$ (3.5pN) for
nonspinning systems, and to lower order for spinning systems.  The
tidal heating $\dot{M}$ is equivalent to a 2.5pN term in the flux
expression.  The full expressions for $E$, $\mathcal{F}$, and
$\dot{M}$ are calculated in
references~\cite{Blanchet:2002,Blanchet:2005a,Blanchet:2005b,Arun:2009,Blanchet:2006gy,Blanchet:2007,Blanchet:2010},
with a collection of errata in reference~\cite{Brown:2007jx}, and are
given explicitly in the accompanying Mathematica notebook
\texttt{PNOrbitalPhase.nb}, which can be found in the ancillary
material with this paper online.  For all the results in this paper,
the full expressions were used.

To integrate the system of ordinary differential equations, various
methods have been developed, each of which should be equivalent at the
level of accuracy of the pN approximation.  These ``approximants''
have been given the names TaylorT1 through
TaylorT4~\cite{DamourEtAl:2001,Buonanno:2006ui,Boyle:2007ft}.  The
TaylorT1 approximant is constructed by directly evaluating the
expressions in Eq.~\eref{eq:PNODEs1}, which are then integrated
numerically.  For the TaylorT4 approximant, the right-hand side of the
expression for $\rmd v / \rmd t$ is first re-expanded in a Taylor
series, and truncated at the appropriate 3.5pN order, which is then
integrated numerically.  The difference between these approximants is
at the level of the 4pN uncertainty.

Alternatively, we can find analytical formulas, rather than
integrating numerically.  The TaylorT2 approximant is derived by
taking the (multiplicative) inverse of the first expression in
Eq.~\eref{eq:PNODEs1} to construct a new system:
\begin{equation}
  \label{eq:PNODEs2}
  \frac{\rmd t}{\rmd v} = - \frac{E'(v)} {\mathcal{F}(v) + \dot{M}(v)}
  \qquad \frac{\rmd\Phi} {\rmd v} = \frac{\rmd\Phi}{\rmd t} \frac{\rmd
    t} {\rmd v} = - \frac{v^{3}}{M}\, \frac{E'(v)} {\mathcal{F}(v) +
    \dot{M}(v)}.
\end{equation}
We re-expand the right-hand sides of these expressions in Taylor
series, truncate at the appropriate 3.5pN order, and integrate with
respect to $v$, obtaining expressions for $t(v)$ and $\Phi(v)$.  This
parametrically determines the phase of the binary as a function of
time, with $v$ being the independent parameter.  Finally, the TaylorT3
approximant is constructed by inverting the series $t(v)$ to obtain
$v(t)$.  There is a 3pN logarithmic term in $t(v)$, which must be
treated as a constant in order to invert the series.  Once this is
done, the series for $v(t)$ can be inserted into $\rmd \Phi / \rmd t =
v^{3}/ M$, which can be integrated analytically to find $\Phi(t)$.

Using any of these approximants, the resulting orbital phase and
frequency allow us to calculate the metric perturbation function $h$.
The perturbation falls off at leading order as $1/r$ and varies with
angle, typically being decomposed in spin-weight $s=-2$ spherical
harmonics~\cite{Brown:2007jx}.  The dominant mode of this
decomposition is generally the $(\ell, m) = (2,2)$ mode.  At leading
order we have\footnote{Note that most references discuss a change of
  variables given by $\Psi \define \Phi - 6\, v^{3} \ln(v/v_{0})$,
  which is a relative 4pN modification to the phase (and can therefore
  be ignored at the level of accuracy currently known for the phase
  evolution).  Here, $v_{0}$ is a freely specifiable parameter related
  to the origin of the time coordinate.  This modification removes
  related terms in expressions for the waveform amplitude by shifting
  them to terms at 4.5pN order and higher, which are currently
  unknown.  We emphasize that---at the level of current pN
  knowledge---this new variable $\Psi$ never needs to be calculated
  explicitly from the known orbital phase and frequency.  Where the
  expressions for $h$ refer to $\Psi$, the standard orbital phase
  $\Phi$ derived from the TaylorT$n$ approximants may be used directly
  in place of $\Psi$ without subtracting the logarithm, and any term
  involving $\ln(v/v_{0})$ should be removed from the expressions for
  amplitude.  %
  %
  %
}
\begin{equation}
  \label{eq:PNAmplitude}
  h_{2,2}(v, \Phi) = \frac{M}{r}\, \sqrt{\frac{\pi}{5}}\, 8\, \eta\,
  v^{2}\, \rme ^{-\rmi 2 \Phi},
\end{equation}
where higher-order terms include additional factors of $v$.  The
additional terms are currently known up to $v^{6}$ (3pN) for
nonspinning systems~\cite{Blanchet:2008}, and to lower order for
spinning systems~\cite{WillWiseman:1996,Brown:2007jx}.  The full
expressions used for this paper, including other modes, are given in
the accompanying Mathematica notebook \texttt{PNWaveform.nb} found in
the ancillary materials with this paper online.

The approximants just discussed describe the pN waveform in the time
domain.  For many purposes, it can be useful to have expressions in
the frequency domain.  The frequency-domain approximant TaylorF2 is
obtained from TaylorT2 using the stationary-phase approximation
(SPA)~\cite{CutlerFlanagan:1994}, together with the approximation that
the gravitational-wave frequency is just twice the orbital frequency,
so $f = v^{3}/\pi\,M$.  Then the frequency-domain waveform is given by
\begin{equation} \fl
  \label{eq:SPA}
  \tilde{h}_{\ell,m}(f) = h_{\ell,m}(v, 0)\, \sqrt{ \frac{2\pi M} {3 m
      v^{2}\, \dot{v}} }\, \exp \left\{\rmi \left[ 2 v^{3}\, t(v)/M -
      m\Phi(v) -\pi/4 \right] \right\},
\end{equation}
where $\dot{v}$ is given in equation~\eref{eq:PNODEs1}, and $t(v)$ and
$\Phi(v)$ are the results of the TaylorT2 approximant.


\section*{References}
\bibliography{../../bibliography/ninja}

\end{document}